\documentclass[reprint,aps,pra,showpacs,superscriptaddress,longbibliography]{revtex4-1}

\usepackage{graphicx}
\usepackage{dcolumn}
\usepackage{bm}
\usepackage{float}
\usepackage{physics}
\usepackage{bm,amsmath}
\usepackage{natbib}
\usepackage{xr}
\usepackage{xcolor}
\usepackage{comment}

\newcommand{\red}[1]{\textcolor{black}{#1}}

\begin{document}

\title{Longitudinal coupling between a Si/SiGe quantum dot and an off-chip TiN resonator}

\author{J. Corrigan}
\thanks{These two authors contributed equally to this work. J. Corrigan present address: Intel Components Research, Intel Corporation, Hillsboro, OR 97124, USA.}
\affiliation{Department of Physics, University of Wisconsin, Madison, WI 53706}
\author{Benjamin Harpt}%
\thanks{These two authors contributed equally to this work. J. Corrigan present address: Intel Components Research, Intel Corporation, Hillsboro, OR 97124, USA.}
\affiliation{Department of Physics, University of Wisconsin, Madison, WI 53706}
\author{Nathan Holman}
\thanks{Present address: HRL Laboratories, LLC., 3011 Malibu Canyon Road, Malibu, CA 90265, USA.}
\affiliation{Department of Physics, University of Wisconsin, Madison, WI 53706}
\author{Rusko Ruskov}
\affiliation{Laboratory for Physical Sciences, 8050 Greenmead Dr., College Park, MD 20740}
\author{Piotr Marciniec}
\affiliation{Department of Physics, University of Wisconsin, Madison, WI 53706}
\author{D. Rosenberg}
\affiliation{MIT Lincoln Laboratory, 244 Wood Street, Lexington, MA 02421}
\author{D. Yost}
\affiliation{MIT Lincoln Laboratory, 244 Wood Street, Lexington, MA 02421}
\author{R. Das}
\affiliation{MIT Lincoln Laboratory, 244 Wood Street, Lexington, MA 02421}
\author{William D. Oliver}
\affiliation{MIT Lincoln Laboratory, 244 Wood Street, Lexington, MA 02421}
\author{R. McDermott}
\affiliation{Department of Physics, University of Wisconsin, Madison, WI 53706}
\author{Charles Tahan}
\affiliation{Laboratory for Physical Sciences, 8050 Greenmead Dr., College Park, MD 20740}
\author{Mark Friesen}
\affiliation{Department of Physics, University of Wisconsin, Madison, WI 53706}
\author{M. A. Eriksson}
\affiliation{Department of Physics, University of Wisconsin, Madison, WI 53706}
\email{maeriksson@wisc.edu}

\date{\today}

\begin{abstract}
Superconducting cavities have emerged as a key tool for measuring the spin states of quantum dots. 
So far however, few experiments have explored longitudinal couplings between dots and cavities, and no solid-state qubit experiments have explicitly probed the ``adiabatic'' regime, where the Purcell decay is strongly suppressed.
Here, we report measurements of a \red{double-quantum-dot charge qubit} coupled to a high-impedance resonator via a ``flip-chip” design geometry. 
By applying an \red{adiabatic} ac drive to the qubit through two different channels, and studying the effects of qubit energy detuning, \red{interdot tunneling, and driving strength,}
we are able to unequivocally confirm the presence of a longitudinal coupling between the qubit and cavity\red{, while the qubit remains in its ground state}.
Since this coupling is proportional to the driving amplitude, and is therefore switchable, it has the potential to become a powerful new tool in qubit experiments.
\end{abstract}

\maketitle

Quantum couplings between microwave resonators and quantum dot qubits are expected to form key components in scalable quantum computing architectures~\cite{Majer:2007p443,Nigg:2017p147701}, with potential applications for both readout and couplings between distant qubits.
Recent experiments have demonstrated spin-photon interactions~\cite{Petersson:2012p380,Mi:2017p156,Landig:2018p179, Samkharadze:2018p1123, Holman:2021p137} and photon-mediated spin-spin interactions \cite{Borjans:2020p195}. 
Most of these have focused on ``transverse'' qubit-cavity couplings, which have some disadvantages, such as the required resonance between the  qubit and cavity~\cite{Burkard:2020p129}, which can be difficult to engineer and which causes rapid decoherence due to the Purcell effect~\cite{Purcell:1946p681,Bienfait:2016p74, Kjaergaard:2020p369}.
Other types of coupling are therefore also of interest, including dispersive and dynamic longitudinal couplings. The latter are expected to be strong and tunable, enabling fast gates and quantum-nondemolition (QND) measurements~\cite{Kerman:2013p123011, Didier:2015p203601,Beaudoin:2016p464003, Royer:2017, Ruskov:2019p245306, Ruskov:2021p035301}. 
Both static \cite{Eichler:2018p227702, Dassonneville:2020p11045, Hazra:2020p152601, Toida:2020p94502} and dynamic longitudinal couplings \cite{Touzard:2019p080502, Ikonen:2019p080503} have been demonstrated experimentally in the superconducting qubit community. 
Longitudinal coupling schemes have also been proposed for quantum-dot qubits, and although initial experiments appear promising~\cite{Bottcher:2022p4773}, their full potential is not yet known.

\begin{figure}[b]
\includegraphics[width=2.8in]{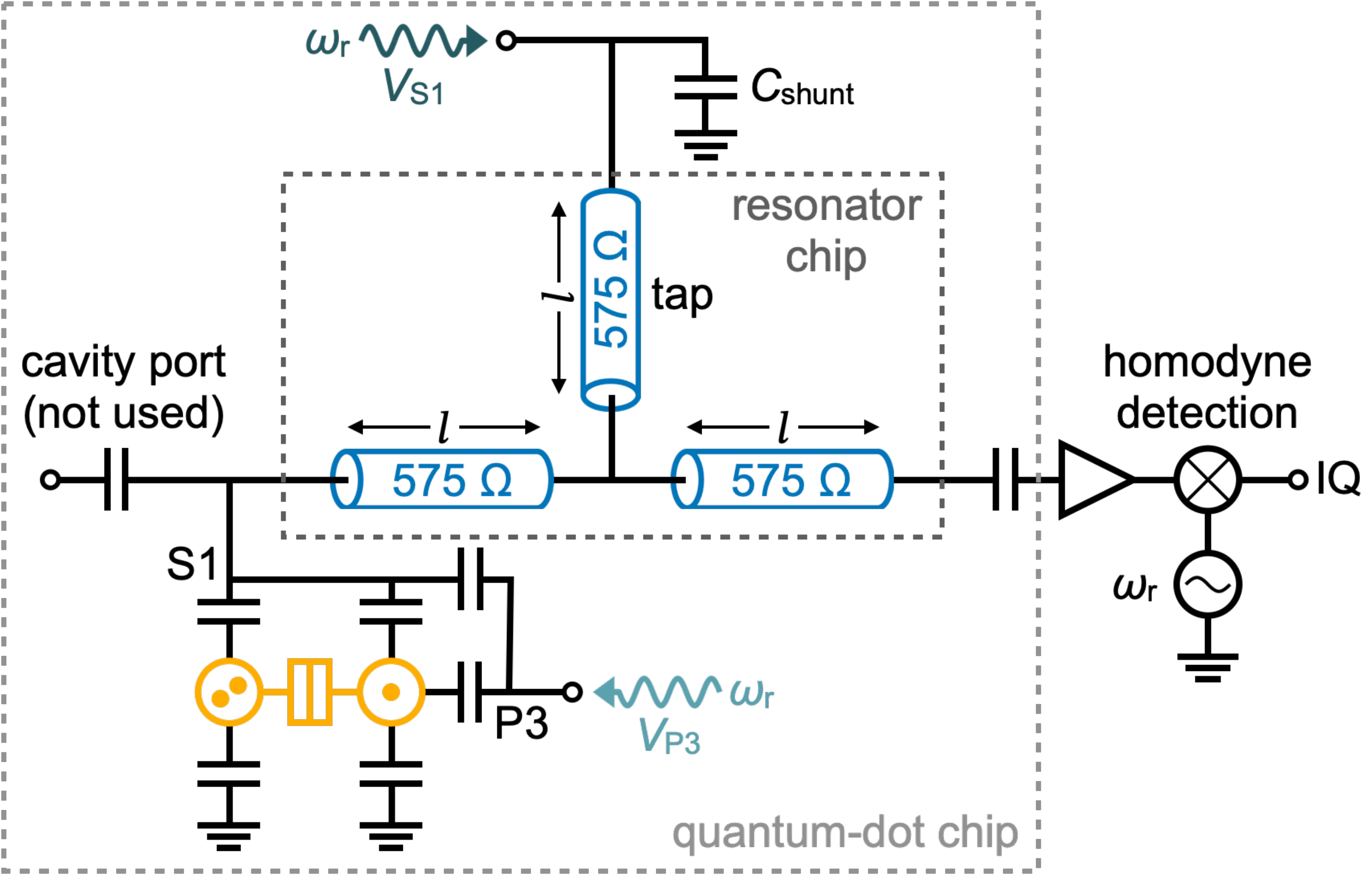}
  \caption{ 
Simplified circuit model for the qubit-cavity device. 
A Si/SiGe double quantum dot (yellow) is capacitively coupled to a high-impedance microwave resonator cavity through gate Sl. 
\red{(The circuit element between the two dots represents a tunnel coupling in parallel with a capacitor.)}
The cavity is made from three TiN coplanar waveguide segments (blue) of length $l=2.02$~mm and characteristic impedance $Z_{0,r}\approx575$~$\Omega$. 
One of the segments is a voltage-node tap with large shunt capacitance $C_\text{shunt}\approx 100$~pF, used for voltage biasing Sl. 
The resonator is vertically integrated on a separate chip from the quantum dots, as indicated (see also Fig.~2). 
During the experiments described below, a microwave tone at cavity frequency $\omega_r/2\pi=1.304$~GHz is applied \red{through non-standard cavity ports}, to either Sl (via the tap segment) or plunger gate P3. 
Transmission through the cavity is measured using a homodyne detection scheme. 
For further details of the device, see Fig.~5 in Appendix~A.}
\label{fgr:couplingterms}
\end{figure}

\begin{figure*}[t]
  \includegraphics[trim=0 60 0 0, clip, width=0.9\textwidth]{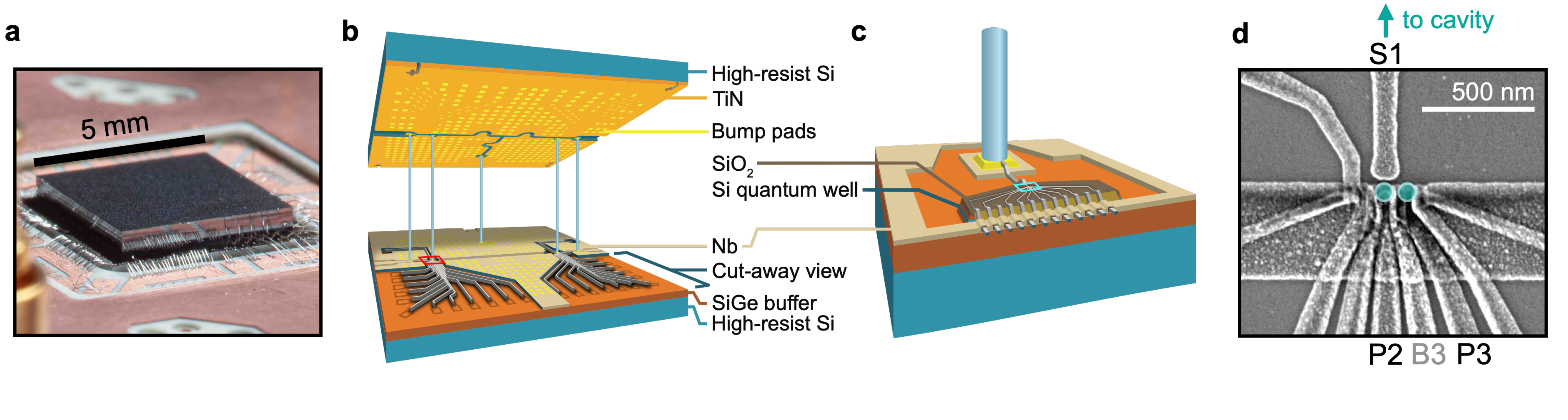}
  \caption{Vertical integration of a quantum-dot qubit and an off-chip microwave cavity.
(a) Image of a connected and wirebonded qubit-cavity flip-chip device. 
(b) Schematic of a 3D-integrated sample, where a high-impedance TiN resonator die is galvanically connected to a Si/SiGe quantum-dot die.  
(c) Blow-up of the bump-bond pad near a double-dot mesa. This region is enclosed by the red rectangle in (b).
(d) Scanning electron micrograph of a double quantum dot nominally identical to the one used in these experiments.
The device is enclosed in the blue rectangle in (c). 
The cavity is driven by high-frequency lines connected to gates P3 or S1.
Plunger gate P2 and barrier gate B3 also are indicated.}
\label{fgr:SEM}
\end{figure*}

In this paper, we report significant enhancement of transmission through an off-chip microwave cavity, arising from the longitudinal coupling to an ac-driven double quantum dot. 
The device is tuned to a charge degeneracy point and driven at the cavity resonance frequency \red{through non-standard cavity ports.} 
The crucial signature of the longitudinal coupling mechanism is its change of sign in the resonator response, depending on which of the two dots is driven. 
The sign change also provides the first direct proof of longitudinal coupling.
We further theoretically describe the longitudinal coupling and its effect on cavity transmission, for a range of 
\red{detuning, tunnel coupling, and ac driving parameters,} obtaining good agreement with the experimental results. 

The sample used in this experiment is a 3D-integrated, high-impedance TiN 
coplanar waveguide resonator on a Si chip coupled to a Si/SiGe quantum-dot chip, as depicted schematically in Fig.~1, pictorially in Fig.~2, and explained in further detail in Ref.~\cite{Holman:2021p137}. 
The 3D nature of the device relieves both wiring and fabrication constraints. 
Wiring is simplified by allowing more space on the qubit die for qubit gate lines and fanout; the low-impedance gates reduce cavity losses~\cite{Holman:2020p083502}. 
Fabrication is simplified because the qubit and cavity chips are no longer constrained to follow the same lithographic processes.  
Figure~\ref{fgr:SEM}(d) shows a scanning electron micrograph of a device nominally identical to the one used in the experiment, in which the quantum dots are capacitively coupled to the resonator through a tap segment on the resonator chip, which is
shunted by a large parallel plate capacitor on the quantum-dot chip. Galvanic contacts between the dies are achieved using underbump metal pads [Fig.~2(c)] formed at the voltage antinodes near the quantum dots, and at the end of the tap segment.
SPICE simulations indicate  that the fundamental resonant mode used in this work, with frequency $\omega_r/2\pi=1.304$~GHz, has weight at both these antinodes and in the tap segment.

\begin{figure}[t]
  \includegraphics[trim=0 0 0 0, clip, width=2.8in]{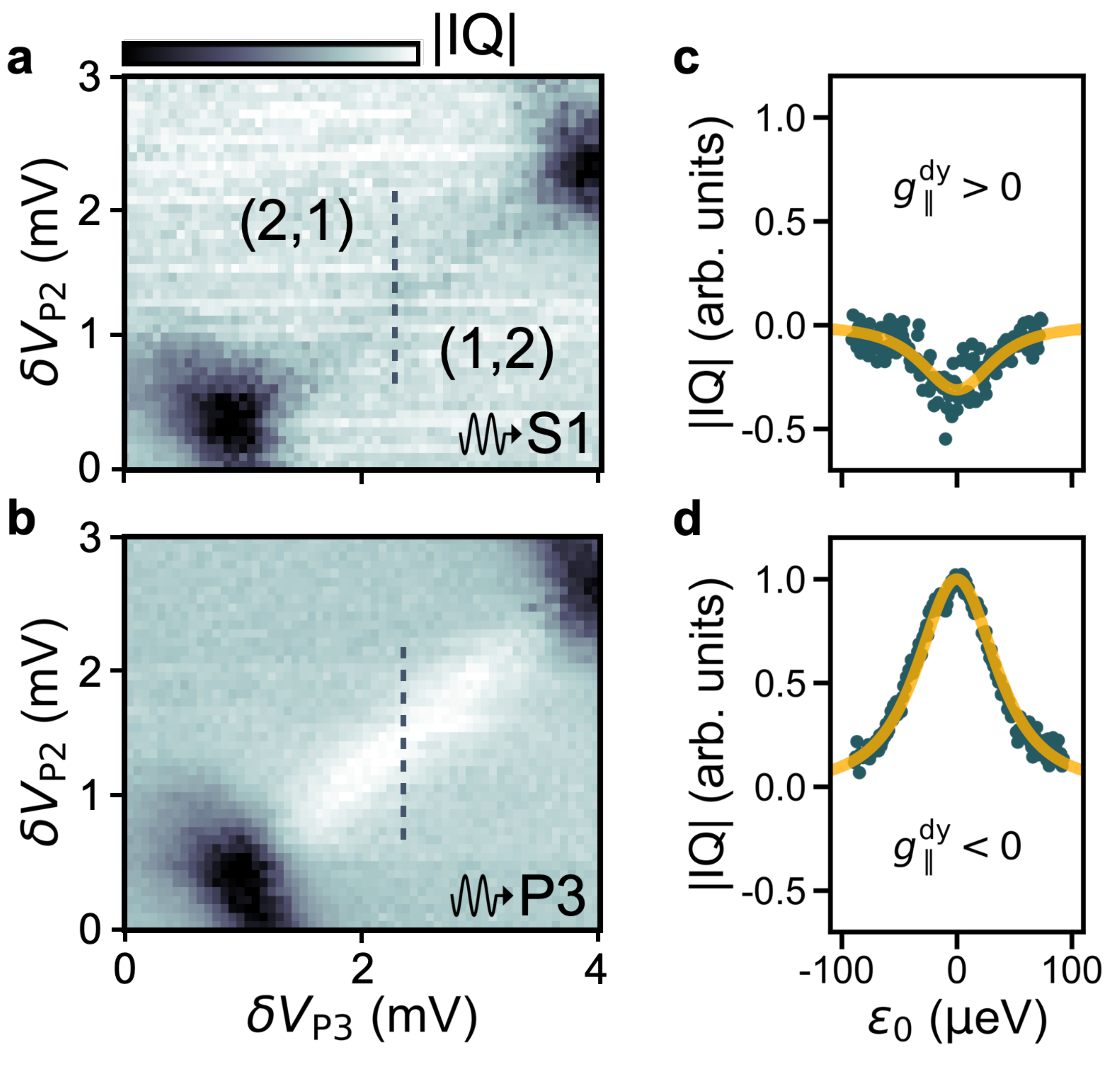}
  \caption{Double-dot stability diagrams, as measured via an off-chip microwave cavity.
(a), (b) IQ transmission measurements of a double dot tuned near its (2,1)-(1,2) charge-occupation transition. The double-dot detuning parameter $\varepsilon(t)=\varepsilon_0+\delta\varepsilon(t)$ is driven by modulating the voltages on gates (a) S1 or (b) P3, with no other changes in device tuning. 
(c), (d) Line cuts across the interdot charging transitions, indicated by the dashed lines in (a), (b). Fits are performed using Eq.~(\ref{eq:IQm}) (solid-yellow curves), following the procedure described in Appendix~C1. Arbitrary IQ units have been rescaled, as described in Appendix~C2, using the same scaling factor in (a)-(d).}
\label{fgr:stability}
\end{figure}

Figures~\ref{fgr:stability}(a) and \ref{fgr:stability}(b) show interdot charging transitions of a three-electron double dot, formed under plunger gates P2 and P3, tuned to its (2,1) or (1,2) charge occupations, as labelled. The data are obtained as a function of the dc voltages applied to gates P2 and P3, while simultaneously ac-modulating the voltages on gates S1 [Fig.~\ref{fgr:stability}(a)] or P3 [Fig.~\ref{fgr:stability}(b)] at the cavity resonant frequency.  
As explained in Appendix~B, the recorded IQ signal is proportional to the microwave signal transmitted through the cavity, which is measured via a homodyne detection scheme.
\red{A dc offset in the homodyne IQ output, which was not calibrated away during experiments, has been subtracted from all data presented in this work.}

The resonator response observed in Fig.~\ref{fgr:stability}(a) is primarily achieved through a longitudinal curvature coupling ($g_\|^\text{dy}>0$) induced by driving the qubit at the cavity resonant frequency. 
A line cut through the charging transition, shown in Fig.~\ref{fgr:stability}(c), provides a clearer view of the suppressed transmission amplitude as a function of the double-dot detuning $\varepsilon_0$, as computed using the measured lever arm $\alpha_\mathrm{P2,\varepsilon}=0.11$~eV/V (see Appendix~D1). 
A dispersive shift ($\delta\omega$) of the cavity resonance frequency, arising from the energy curvature of the double-dot energy levels, further contributes to suppression of the signal.
However, through the analysis below, we find its effect to be much weaker than that of the longitudinal coupling. Note here that by adiabatically sweeping the gate voltages in this stability diagram, we ensure that the qubit remains in its ground state at all times. 
Thus here, and throughout this work, it is the magnitude of qubit-cavity coupling, not the qubit state itself, that changes near the charging transition. 

The behavior observed in Fig.~\ref{fgr:stability}(b), and its corresponding line cut in Fig.~\ref{fgr:stability}(d) is extraordinary for being opposite of the conventional behavior, and clearly inconsistent with a dispersive frequency shift. 
Instead, it indicates an enhancement of the stationary field in the resonator. Remarkably, the observed qubit-cavity coupling ($g_\|^\text{dy}<0$) not only has the opposite sign as the response in Fig.~\ref{fgr:stability}(a), it is also much larger in magnitude and provides a much cleaner signal.
This is because P3’s detuning lever arm $\alpha_\mathrm{P3,\varepsilon}=0.09$~eV/V is more than twice as large as S1’s, $\alpha_\mathrm{S1,\varepsilon}=0.04$~eV/V. 
Therefore, modulating the voltage on P3 applies a stronger drive to the qubit, which increases the magnitude of the coupling and the measured resonator response. 
Aside from our choice of driving gate, there are no other electrostatic changes in the double-dot detuning or tunnel coupling between Figs.~\ref{fgr:stability}(a) and \ref{fgr:stability}(b). 
Hence, the heightened response in Fig.~\ref{fgr:stability}(b)~\footnote{The different background noise levels observed in the experimental configurations of Figs.~3(a) and 3(b) can be partially explained by the different numbers of microwave photons in the resonator (see Appendix~E2), since the noise level scales as $1/\sqrt{\langle n\rangle }$~\cite{MacDonaldBook}}, \red{which 
we show below to be} proportional to the amplitude of the qubit drive, has immediate and important consequences for qubit readout. 

To understand these different effects, we construct a Hamiltonian for our qubit system.
As argued in Appendix~A, since we work in an operating regime very near the charging transition, $\varepsilon_0=0$, and since the data in Figs.~\ref{fgr:stability}(c) and \ref{fgr:stability}(d) appear symmetric about $\varepsilon_0= 0$, it should be a good approximation to ignore other nearby level crossings.
We therefore treat the system as a simple charge qubit, in which an excess electron moves between two sides of a double dot, atop a fixed (1,1) charge configuration.
The resulting qubit Hamiltonian is given by $H_q=(\varepsilon/2)\sigma_z+t_c\sigma_x$, where the double-dot detuning parameter $\varepsilon=\varepsilon_0+\varepsilon_q\text{cos}(\omega t)$ includes both static and driving terms, $t_c$ is the tunnel coupling between the two dots, and $\sigma_x$ and $\sigma_z$ are Pauli operators acting on the charge basis of left or right-localized states.
Since $\sigma_z$ is proportional to the charge-dipole operator of the double dot and $(a+a^\dagger)$ corresponds to the quantum field of the resonator ($a^\dagger$ and $a$ are photon creation and annihilation operators), the full Hamiltonian of the coupled qubit-cavity system is given by $H=H_q+\hbar\omega_ra^\dagger a+2\varepsilon_r\cos(\omega_rt)(a+a^\dagger)+\hbar g_0\sigma_z(a+a^\dagger)$, where $g_0$ is the bare coupling between the resonator and double dot, and we have included a resonant cavity-driving term, with amplitude $2\varepsilon_r$.
In the absence of coupling ($g_0=0$) or qubit driving ($\varepsilon_q=0$), the bare qubit energy is given by $E_{q0}=\hbar\omega_{q0}=\sqrt{\varepsilon_0^2+4t_c^2}$.

In all of the experiments reported here, the device is operated in the adiabatic coupling regime $\omega_r\ll\omega_{q0}$, in which the resonator is unable to excite the qubit.
We further consider the weak-driving, dispersive limit, $\varepsilon_q,\hbar g_0\ll t_c$.
As explained in Appendix~A, an effective Hamiltonian can be derived under these assumptions.
Moving to a frame rotating at the cavity frequency, performing a rotating-wave approximation, and retaining coupling terms up to order ${\cal O}[\varepsilon_qg_0,g_0^2]$, the system Hamiltonian reduces to
\begin{multline}
H_\rho \approx \frac{E_{q0}}{2}\tilde\sigma_z 
+\varepsilon_r(\tilde a+\tilde a^\dagger)
 \\
 +\frac{\hbar g_\|^\text{dy}}{2}\tilde\sigma_z(\tilde a+\tilde a^\dagger)
+\hbar\delta\omega\,\tilde \sigma_z\left(\tilde a^\dagger \tilde a+\frac{1}{2}\right), 
\end{multline}
where the tildes on the operators $\tilde\sigma_z$, $\tilde a$, and $\tilde a^\dagger$ indicate that they are defined in the rotating frame of the qubit energy-basis states (see Appendix~A).
The effective coupling terms are given by
\begin{equation}
g_\|^\text{dy}=4t_c^2g_0\varepsilon_q/E_{q0}^3
\quad \text{and} \quad
\delta\omega=8\hbar t_c^2g_0^2/E_{q0}^3 .
\label{eq:params}
\end{equation}
Such couplings can be observed as effective shifts in the cavity field or cavity energy that depend on the state of the qubit. 
They are known as energy-curvature or quantum-capacitance couplings~\cite{Ruskov:2019p245306, Ruskov:2021p035301}, and we refer to $g_\|^\text{dy}$ and $\delta\omega$ here as the dynamic longitudinal and dispersive couplings, respectively.
It can be seen that $\delta\omega$ differs from the conventional definition~\cite{Blais2021p025005} of the dispersive coupling ($\chi=g_\perp^2/\omega_q=\delta\omega/2$) by a factor of 2, due to our experiment being performed in the ultra-dispersive or ``adiabatic'' regime, $\omega_r\ll \omega_q$ (see Appendix~A).

Although the microwave resonator cannot excite the qubit in the present scheme, the qubit has a measurable effect on the resonator through $\delta\omega$ and $g_\|^\text{dy}$, with the latter being proportional to the ac drive applied to the qubit detuning parameter, $\varepsilon_q$.
It is crucial to note that an ac drive applied through gate S1 in Fig.~\ref{fgr:SEM}(d) primarily affects the left dot, while an ac drive on gate P3 primarily affects the right dot.
The two drives therefore act on $\varepsilon_q$ with opposite sign.
On the other hand, the two driving channels act on the cavity (and therefore $\varepsilon_r$) with the same sign.
Here, S1 drives the cavity directly, while P3 drives the cavity indirectly through crosstalk between the two gates.
As explained in Appendix~A, this difference in signs is crucial for observing opposing behaviors in Figs.~{\ref{fgr:stability}}(a) and {\ref{fgr:stability}}(b).
Indeed, for future qubit readout experiments, it could be preferable to drive the qubit on the same side as the resonator coupling (i.e., through P2 rather than P3), so that $\delta\omega$ and $g_\|^\text{dy}$ affect the resonator in tandem, rather than opposition, to achieve a stronger total effect.

Finally, we provide an expression for the output of our homodyne detector (see Appendix~B), which  converts the signal transmitted through the microwave cavity into a dc voltage:
\begin{equation}
\text{IQ}_-
=c\frac{(\varepsilon_r/\hbar)-(g_\|^\text{dy}/2)}{\sqrt{\delta\omega^2+\kappa^2/4}} . \label{eq:IQm}
\end{equation}
Here, the minus sign in the subscript indicates that the qubit is always in its ground state, $c$ is a proportionality constant, relating the signal transmitted through the microwave cavity to the output voltage from the IQ mixer, and $\kappa/2\pi$ is the photon loss rate from the cavity.

Equations~(1)-(3) are used to quantitatively fit the data in Figs.~\ref{fgr:stability}(c) and \ref{fgr:stability}(d), to determine the different coupling parameters.
The fitting procedure is explained in detail in Appendix~C1, and the results are shown here as yellow curves.
After fitting, we subtract the asymptotic background signal in IQ$_-$, which varies around 10\% between the two drive configurations.
Note that the proportionality constant $c$ in Eq.~(3) is related to the measurement setup rather than the qubit-cavity coupling, and cannot be measured directly.
Since $\varepsilon_r$ and $g_\|^\text{dy}$ are both multiplied by $c$, their values cannot be determined independently.
However, the tunnel coupling $t_c$ can be obtained without ambiguity from the fitting procedure, and the bare qubit-cavity coupling $g_0/2\pi=5.5$~MHz can be determined from measured device parameters Appendix~D2.
Combining these results with knowledge of the applied drive power allows us to estimate $\varepsilon_r$ and $g_\|^\text{dy}$ in the current experiments, as described below.

\begin{figure}[t]
  \includegraphics[trim = 0 0 0 0, clip, width = 3.1in]{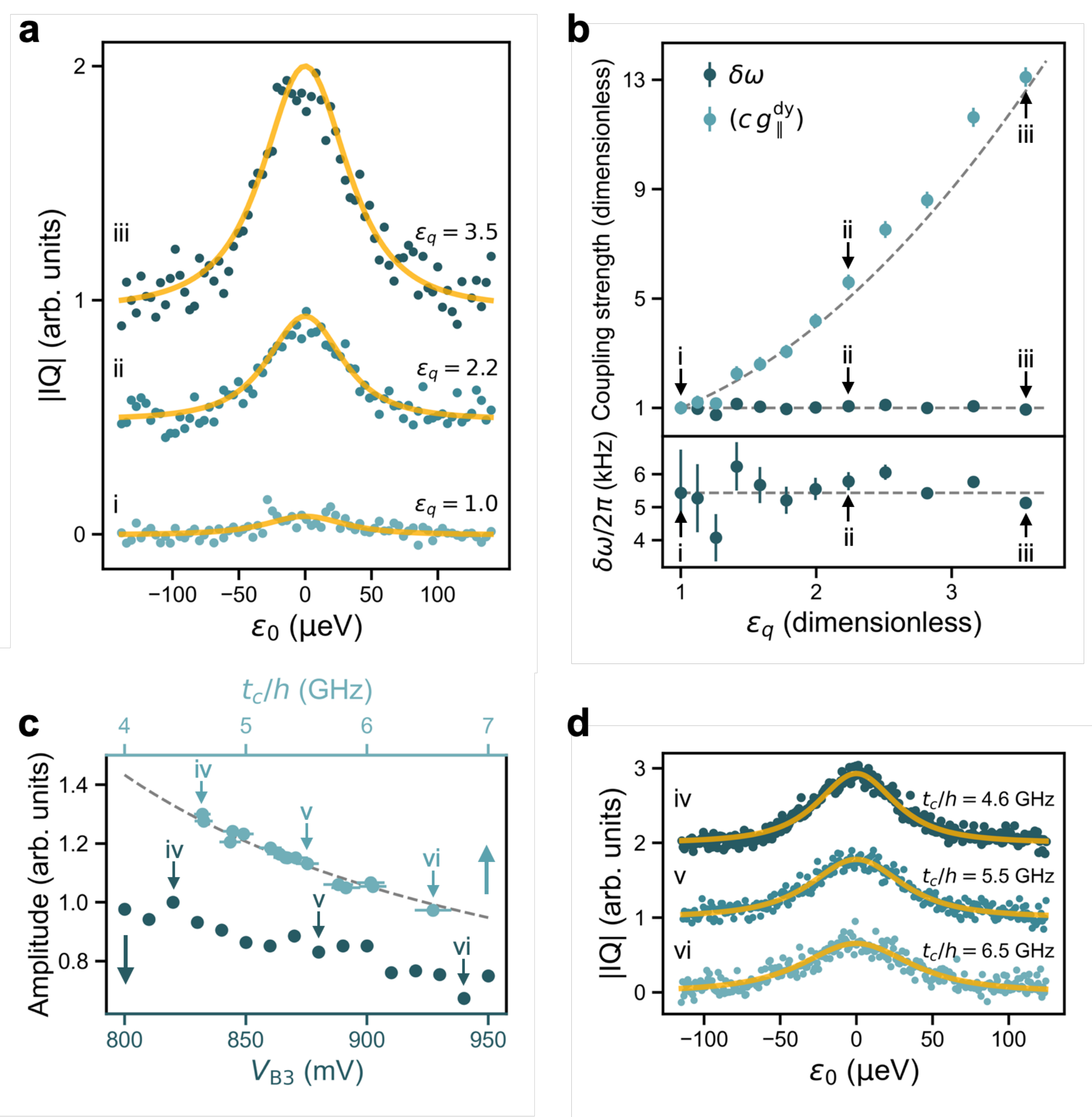}
  \caption{
\red{Dependence of the cavity transmission and cavity-qubit couplings on ac driving amplitude and tunnel coupling,} for P3 driving near the interdot charging transition.
\red{(a) IQ transmission traces for three different values of the ac driving strength (dimensionless units, see main text): (i) $ \varepsilon_q = 1.0$, (ii) 2.2, and (iii) 3.5. 
Data are normalized to the height of the tallest peak [(iii)], and the traces are offset vertically by 0.5 for clarity, after subtracting off background levels. 
Yellow curves are fits to Eq.~(\ref{eq:IQm}).}
\red{(b) Upper plot: fitting values for the product $cg_\|^\text{dy}$ and for $\delta\omega$, evaluated at the peak value $\varepsilon_0=0$, as a function of $\varepsilon_q$ (also dimensionless). 
Lower plot: same $\delta\omega$ results, plotted in absolute units. 
In both panels, the dashed gray curves correspond to expected scaling behavior: no $\varepsilon_q$ dependence for $\delta\omega$, and parabolic dependence for $cg_\|^\text{dy}$. 
Arrows indicate the traces shown in (a).}
(c) Dark-teal data points: transmission peak amplitudes vs.\ tunnel-barrier gate voltage (bottom axis).
Background levels have been subtracted and normalized to the height of the tallest peak.
Light-teal data points: peak amplitude vs.\ tunnel coupling $t_c$ (top axis), determined from transmission peak fits. 
These data points have been offset by 0.3 arbitrary units for clarity. 
The amplitudes follow a $1/t_c$ trend (gray-dashed curve). 
\red{(Note that the top and bottom axes are not linearly related, as the dependence of $t_c$ on $V_\text{B3}$ is generally exponential.)}
(d) IQ transmission traces for the voltage values indicated by arrows in (c): (iv)~$V_\text{B3}=820$, (v) 880, or (vi) 940~mV. The data are fit to Eq.~(\ref{eq:IQm}) (yellow curves), yielding $t_{c}$ values 4.6, 5.5, and 6.5~GHz, respectively. For clarity, each of the traces is offset by 1 in these arbitrary units.}
\label{fgr:tunnelcoupling}
\end{figure}


We now further explore the physics of $g_\|^\text{dy}$ by \red{varying different device parameters that affect the coupling strength.
(For all remaining measurements reported here, the ac drive is applied to gate~P3.)
Figure~\ref{fgr:tunnelcoupling}(a) shows $\varepsilon_0$ line cuts labeled (i)-(iii), for three different ac driving amplitudes $\varepsilon_q$.
The IQ units for this plot are normalized to the height of the tallest peak [(iii)], and the plots are shifted vertically for clarity.
The reported $\varepsilon_q$ units are also relative, since we cannot determine this parameter precisely.
We see that the transmission peaks increase  significantly in height with $\varepsilon_q$.
For comparison, we note that these data were acquired at a similar device tuning as Fig.~\ref{fgr:stability}, where the relative $\varepsilon_q$ value was 3.5.}

\red{
We can extract the quantum couplings $\delta\omega$ and $g_\|^\text{dy}$ at different driving strengths $\varepsilon_q$ by fitting the transmission peaks to Eq.~(\ref{eq:IQm}), following the procedure described in Appendix~C.
Three such fits are shown in yellow in Fig.~\ref{fgr:tunnelcoupling}(a). 
Figure~\ref{fgr:tunnelcoupling}(b) shows the full set of fitting results for $\delta\omega$, and the product parameter $cg_\|^\text{dy}$ (see Appendices~B and C), evaluated at the $\varepsilon_0=0$ transmission peak, as a function of $\varepsilon_q$.
The points corresponding to the three data sets in Fig.~\ref{fgr:tunnelcoupling}(a) are indicated by arrows.
Because we are unable to estimate parameters $c$ and $g_\|^\text{dy}$ independently in this fitting procedure, we scale $\delta\omega$ and $cg_\|^\text{dy}$ by their values at $\varepsilon_q=1$ in the top panel.
(In the bottom panel, $\delta\omega$ is also reported in dimensionful units.)
We note that the $cg_\|^\text{dy}$ results in Fig.~\ref{fgr:tunnelcoupling}(b) follow a parabolic scaling law with respect to $\varepsilon_q$ (dashed gray curve).
Since $c \propto \varepsilon_q$ for our homodyne detection scheme, as explained in Appendix~B, we infer that $g_\|^\text{dy} \propto \varepsilon_q$, as consistent with Eq.~(\ref{eq:params}).
This confirms the parametric nature of $g_\|^\text{dy}$, and has important implications for qubit tunability, as we discuss below.}

\red{
The dispersive-like coupling term $\delta\omega$ is not expected to exhibit $\varepsilon_q$ dependence.
Unfortunately, the peak-fitting procedure described above does not enable a direct measurement of $\delta\omega$ because of its small size compared to $g_\|^\text{dy}$; however it does allow us to determine $t_c$.
In Fig.~\ref{fgr:tunnelcoupling}(b), we therefore plot $\delta\omega$ computed from Eq.~(2).
Alternatively, we can also obtain direct measurements of $\delta\omega$ in a slightly different tuning regime, as described in Appendix~G, obtaining results very similar to Fig.~\ref{fgr:tunnelcoupling}(b).
These combined results indicate that $\delta\omega$ does not depend on the driving strength $\varepsilon_q$, as consistent with our theoretical understanding, and they confirm that the dispersive-like coupling is very small in the ultra-dispersive regime, with $\delta\omega/2\pi\approx 2.6$-6.2~kHz.}

\red{We can also explore the variation of $g_\|^\text{dy}$ with tunnel coupling.
In this case, Fig.~\ref{fgr:tunnelcoupling}(c) indicates that the transmission peak amplitude decreases with $t_c$.}
(Here, we only vary $V_\text{B3}$, 
keeping all other device parameters fixed.)
Note that this experiment was performed on a different date, resulting in different device tunings than Figs.~\ref{fgr:stability}, \ref{fgr:tunnelcoupling}(a), and \ref{fgr:tunnelcoupling}(b), and a different value of $g_0/2\pi = 4.1$~MHz (see Appendix~D2), due to electrostatic drift.

To analyze the data, we first extract the peak heights.
This is done using a Lorentzian fitting form, to avoid assumptions about the physical model; however results are essentially identical to fits using Eq.~(\ref{eq:IQm}).
The Lorentzian peak heights are plotted in Fig.~\ref{fgr:tunnelcoupling}(c) after normalizing the data to the tallest peak, labeled (iv).
We also fit the IQ peaks to Eq.~(\ref{eq:IQm}).
The detailed fitting procedure is described in Appendix~C2.
Three example fits are shown in Fig.~\ref{fgr:tunnelcoupling}(d), corresponding to the three labeled points (iv)-(vi) in Fig.~\ref{fgr:tunnelcoupling}(c).
The peak heights are replotted in Fig.~\ref{fgr:tunnelcoupling}(c) as a function $t_c/h$, as determined through this procedure.
The behavior is consistent with our expectations from Eqs.~(\ref{eq:params}) and (\ref{eq:IQm}), that $\text{peak amplitude}\sim g_\|^\text{dy}\propto 1/t_c$. 

The global parameter $c$ in Eq.~(\ref{eq:IQm}) precludes being able to specifically determine $\varepsilon_r$ or $\varepsilon_q$ in our fitting procedure.
However, by taking into account the full microwave power budget, including input power and all significant attenuation sources, we estimate P3 driving amplitudes as large as $\varepsilon_q/h \approx 152$~MHz in these experiments (see Appendix~C).
Indeed, we expect P3 to provide the largest longitudinal interaction, due to its strong capacitive coupling to the dot.
Using this estimate and fitting results for $t_c$, we use Eq.~(\ref{eq:params}) to estimate $|g_\|^\text{dy}/2\pi|\approx 12$-68~kHz in the current experiments, including Figs.~3 and 4.
Since $g_\|^\text{dy}$ is proportional to $\varepsilon_q$, we emphasize that it should be possible to further enhance the coupling through stronger driving, while still remaining in the weak-driving regime. 
Although we did not explore this possibility here, it remains an important direction for future work. 
To estimate coupling values that could potentially be achieved in the present device, we take $\varepsilon_q / t_c = 0.1$ as an upper bound on weak driving, yielding a maximum value of $|g_\|^\text{dy}/2\pi|=0.28$~MHz, which is much larger than our estimates for $\delta\omega$, described above. 

It is crucial to note that the current experiments are performed in the far-detuned regime, $\omega_{q0}/\omega_r \gtrsim 7$, which we refer to as ``adiabatic'' here, because the resonator cannot excite the qubit. 
To our knowledge, previous longitudinal coupling experiments have never accessed this extreme-detuning regime, either in superconducting~\cite{Touzard:2019p080502, Ikonen:2019p080503} or quantum dot qubit implementations~\cite{Bottcher:2022p4773}.
The adiabatic regime has several appealing features: (1) Purcell decoherence is strongly suppressed, since we are far from resonance; (2) there are no engineering constraints related to satisfying a resonance condition; (3) the coupling can be turned on and off via the driving term $\varepsilon_q$.
Focusing on this final point, we note that when $\varepsilon_q=0$, the total coupling is reduced to the small, residual dispersive coupling $\delta\omega$. 
When larger couplings are desired, they can be achieved by increasing $\varepsilon_q$ up to the upper bound on weak driving.
A useful figure of merit is therefore given by the tunability ratio, $|g_\|^\text{dy}/\delta\omega|=\varepsilon_q/2\hbar g_0$, which can be large when the driving amplitude $\varepsilon_q$ is much larger than \red{the zero-point fluctuations of the cavity, as felt by the dot.}
Estimates for the current device indicate a tunability factor of 18.4.
However, we expect that tunability factors as large as 79 could be achieved \red{for the tunnel couplings measured here, if the drive was increased to the weak-driving limit.}
These estimates differ considerably from recent experiments in GaAs~\cite{Bottcher:2022p4773}, where much larger values of the dispersive coupling $\delta\omega/2\pi\sim 0.1$-0.25~MHz were obtained~\footnote{Here, we compare to the dispersive coupling definition $\chi_B$ used in \cite{Bottcher:2022p4773}, which is unconventional~\cite{Blais2021p025005}, and differs from our definition of $\delta\omega$ by a factor of 2.}, due to the large bare resonator coupling $g_0$.
A larger longitudinal coupling $g_\|^\text{dy}/2\pi \sim 1$~MHz was also observed in \cite{Bottcher:2022p4773}, yielding figures of merit in the range of 1-5, much smaller than those reported here.
These differences emphasize the inherent competition between the strength ($\propto$$g_0$) and the tunability ($\propto$$g_0^{-1}$) of the longitudinal coupling, \red{where $g_0$ is an engineered variable.}

In conclusion, we have explored the energy curvature couplings $\delta\omega$ and $g_\|^\text{dy}$ between a Si/SiGe \red{double-dot charge qubit} and an off-chip, high-impedance microwave resonator. 
By varying the tunnel coupling and applying the drive through different channels \red{and at different strengths}, we are able to prove the existence of the dynamic longitudinal coupling.
We estimate its \red{potential} strength here as $|g_\|^\text{dy}/2\pi|\approx 68$~kHz, which can be increased to 0.28~MHz upon stronger driving, even in the far-detuned regime.
We compare $g_\|^\text{dy}$ to the dispersive coupling $\delta\omega$, finding $|g_\|^\text{dy}/\delta\omega|>18$ here, and potentially up to 79 under optimal conditions.
Moreover such strong coupling \red{($|g_\|^\text{dy}/\delta\omega|\gg 1$)} is achieved here in the ultra-dispersive (``adiabatic'') regime, which has not been previously explored.
\red{Crucially, such strong, dispersive couplings reduce engineering constraints and suppress Purcell decoherence.}
We conclude that such longitudinal driving forms an important tool for future quantum-dot qubit experiments.

The authors thank HRL for support and L.~F.~Edge for providing the Si/SiGe heterostructure used in this work. Research was sponsored in part by the Army Research Office (ARO) under Grant Number W911NF-17-1-0274 and by the Department of Defense. 
JC acknowledges support from the National Science Foundation Graduate Research Fellowship Program under Grant No. DGE-1747503 and the Graduate School and the Office of the Vice Chancellor for Research and Graduate Education at the University of Wisconsin-Madison with funding from the Wisconsin Alumni Research Foundation.
We acknowledge the use of facilities supported by NSF through the UW-Madison MRSEC (DMR-1720415) and the MRI program (DMR–1625348).
Work done at MIT Lincoln Laboratory was funded in part by the Assistant Secretary of Defense for Research \& Engineering under Air Force Contract No. FA8721-05-C-0002. 
The views, conclusions, and recommendations contained in this document are those of the authors and are not necessarily endorsed nor should they be interpreted as representing the official policies, either expressed or implied, of the Army Research Office (ARO) or the U.S. Government. The U.S. Government is authorized to reproduce and distribute reprints for Government purposes notwithstanding any copyright notation herein.




\appendix

\section{Derivation of the Curvature Coupling Terms}

\red{
In this work, we have studied the interactions between an off-chip microwave resonator and a charge qubit. Although we do not explicitly perform qubit operations, the work is motivated by qubit operations, including readout. 
In some cases, the qubit of interest could be a charge qubit, for which the information is stored in the charge degree of freedom.
In other cases, the information is stored as a spin, then converted to charge information to facilitate readout.
In both cases, the simplest system we could consider is comprised of a single electron in a double dot.
However the motivation for the current work lies with experiments performed on the quantum dot hybrid qubit, which can be viewed as a cross between a spin and a charge qubit~\cite{Shi:2012p140503,Kim:2014p70}.
The device incorporates an odd number of three or more electrons in a double dot, and the corresponding gate operations are performed in the vicinity of a charging transition ($\varepsilon\approx 0$), where one of the electrons tunnels between the dots.
Below, we first explain why the following theoretical analysis can be framed in terms of the simpler problem, comprising a double dot with one electron.}

The relevant basis states for the quantum dot hybrid qubit are fourfold, including two charge configurations and a pair of spin-orbit states, with a characteristic energy splitting approximately equal to the singlet-triplet splitting $E_\text{ST}$ of the dot containing an even number of electrons.
When $E_\text{ST}$ is somewhat larger than both the detuning parameter $\varepsilon$ and the inter-dot tunnel coupling $t_c$, we may ignore one or both of the high-energy excited states.
For example, in many cases, it is sufficient to consider a three-level description~\cite{Koh:2012p250503}.
In the present work, we do not exploit the full hybrid-qubit functionality, and we further make use of the fact that the data in Figs.~3 and 4 of the main text appear symmetric with respect to $\varepsilon_0=0$ (as consistent with $\varepsilon_q,t_c<E_\text{ST}$) and do not hint at nearby energy-level anticrossings, thus reinforcing such an approximate two-level description.
We may therefore restrict our analysis to just the two lowest-energy states. The resulting qubit Hamiltonian is then given by $H_q=(\varepsilon/2)\sigma_z+t_c\sigma_x$.

\begin{figure}[t]
\includegraphics[width = 0.4\textwidth]{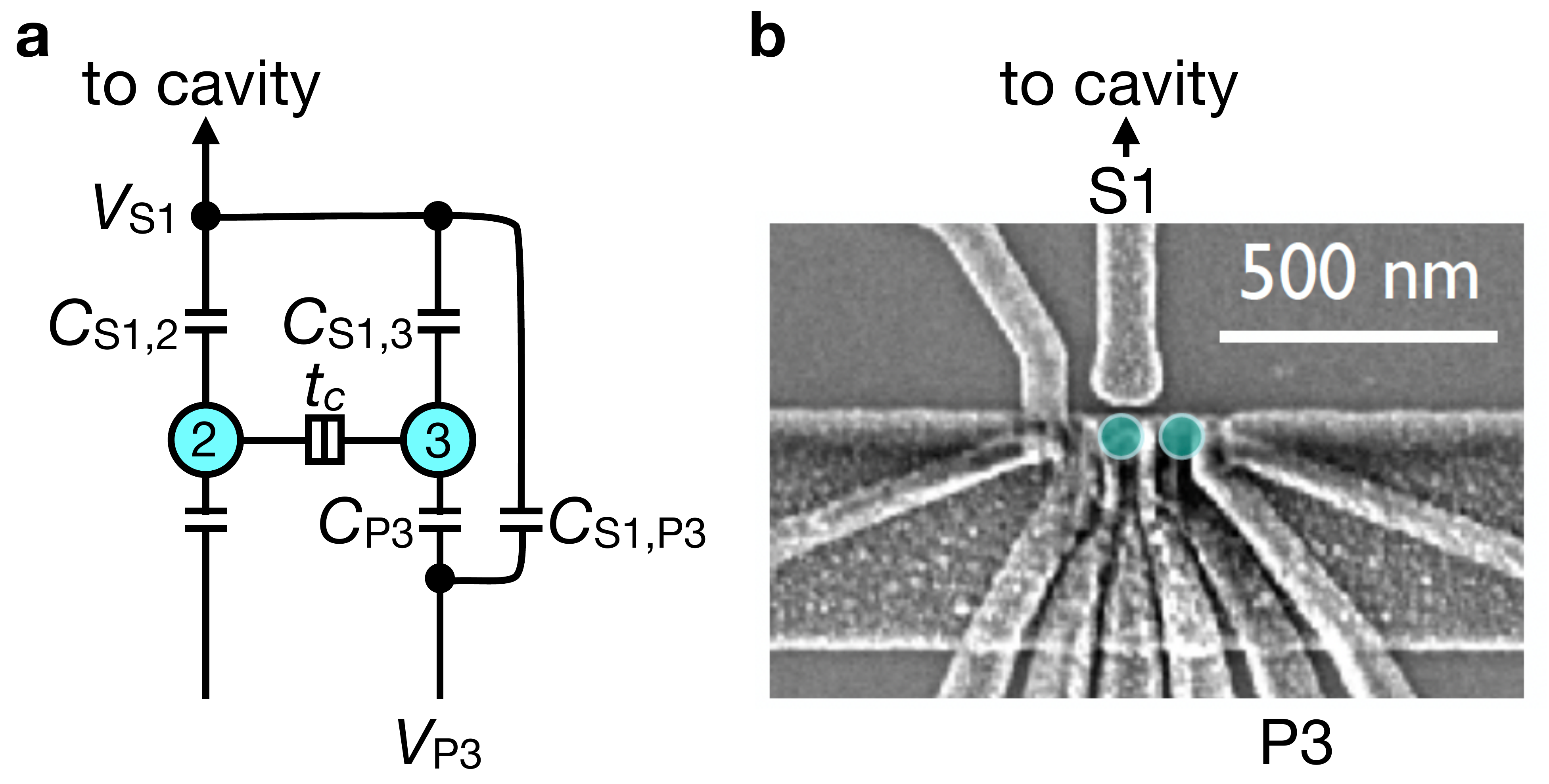}
  \caption{ 
(a) Simplified circuit diagram of a double dot coupled to an external microwave cavity. 
Direct capacitances $C_\text{S1,2}$ and $C_\text{P3}$ between the dots and gates S1 and P3 are indicated.
Cross capacitances $C_\text{S1,3}$ and $C_\text{S1,P3}$ are also indicated, as is the double-dot tunnel coupling $t_c$.
(Note that other capacitances are present in our system, such as couplings to ground, which do not play a direct role in the current discussion and are not shown here.)
The voltage $V_\text{S1}$ on gate~S1 also couples to the cavity. 
Voltages $V_\text{S1}$ and $V_\text{P3}$ may both include an ac drive, although only one gate is driven at a time.
(b) A triple-dot device, operated in double-dot mode, reproduced from Fig.~2(d) of the main text.
In (a), the two activated dots are labeled 2 and 3.}
\label{fig:circuit}
\end{figure}

To construct the full Hamiltonian, we consider the circuit illustrated in Fig.~\ref{fig:circuit}, which indicates the relevant voltages and capacitances in our device, formed of a double dot coupled to a microwave cavity.
As described in the main text, microwave driving tones are applied in two ways to this circuit.

In the first driving mode, an ac voltage is applied to gate~S1, which has two effects on the Hamiltonian.
First, the drive causes a modulation of the double-dot detuning parameter (primarily via dot~2), as described by the Hamiltonian term $(\varepsilon_2/2)\cos(\omega t)\sigma_z$, where $\varepsilon_2$ is the driving amplitude of the detuning parameter via dot~2.
There is also a secondary effect of driving gate~S1, due to the  cross-capacitances labelled $C_\text{S1,3}$ and $C_\text{S1,P3}$ in Fig.~\ref{fig:circuit}(a), which slightly reduces the total detuning drive.
\red{Such cross-capacitances modify the primary capacitive couplings;} however, their effect is \red{not leading order, and therefore} modest, and will be neglected here.
A much more important contribution to the Hamiltonian, associated with S1-driving, is the cavity coupling, which takes the form $2\varepsilon_r\cos(\omega t)(a+a^\dagger)$.
We note that the driving amplitudes $\varepsilon_2$ and $\varepsilon_r$ arise from the same voltage oscillations on gate S1, and are therefore proportional.
Although it is possible to specify this proportionality constant in terms of lever arms, it is convenient to simply define $\varepsilon_r=\beta_2\varepsilon_2$.
In the fitting procedure described below, $\beta_2$ and $\varepsilon_2$ are chosen as fitting parameters, while $\varepsilon_r$ becomes a dependent parameter.

The P3 driving mode also modulates the detuning parameter (primarily via dot~3), as described by the Hamiltonian term $(\varepsilon_3/2)\cos(\omega t)\sigma_z$, where $\varepsilon_3$ is the driving amplitude of the detuning parameter via dot~3.
Note that although we have neglected cross-capacitances in the S1-driving Hamiltonian, it is very important to include them for the case of P3 driving. 
This is because the cross-capacitances are solely responsible for driving the cavity in this scenario.
To account for this, we introduce the proportionality relation $\varepsilon_r=\beta_3\varepsilon_3$.
The cross-capacitance also has a secondary effect on the total detuning through dot~2, of the form $\varepsilon_2=(\beta_3/\beta_2)\varepsilon_3$.
The total driving amplitude for the detuning parameter is therefore slightly reduced to $(1-\beta_3/\beta_2)\varepsilon_3$ due to cross talk.

For the experiments described here, it is crucial to note that the S1 and P3 driving terms contribute to the total detuning with opposite sign, because they are applied on opposite sides of the double dot.
Taking into account both driving terms, the total detuning thus becomes 
\begin{equation}
\varepsilon=\varepsilon_0+\varepsilon_q\cos(\omega t),\quad\quad\text{where}\quad\quad\varepsilon_q=\varepsilon_2-(1-\beta_3/\beta_2)\varepsilon_3. 
\end{equation}
On the other hand, the two driving terms contribute to the cavity drive with the same sign:
\begin{equation}
\varepsilon_r=\beta_2\varepsilon_2+\beta_3\varepsilon_3 .
\end{equation}
In both of these equations, we emphasize that only one of the drives, $\varepsilon_2$ or $\varepsilon_3$, is active at a time.
The full Hamiltonian is finally given by
\begin{multline}
H=\frac{\varepsilon_0
+\varepsilon_q\cos(\omega_rt)}{2}\sigma_z
+t_c\sigma_x
+2\varepsilon_r\cos(\omega_rt)(a+a^\dagger) \\
+\hbar\omega_ra^\dagger a
+\hbar g_0\sigma_z(a +a^\dagger) ,
\end{multline}
where in addition to the terms already discussed, the fourth term describes a microwave cavity with resonant frequency $\omega_r$, and the fifth term describes a coupling of strength $g_0$ between the cavity and the qubit. 
As a note of clarification, the parameter $\beta_2$ is used here to describe the relation between the oscillating chemical potential of dot 2 and the cavity drive, while $g_0$ describes the capacitive coupling between the qubit electron and the cavity.
Also note that we have set all driving frequencies equal to the cavity resonance frequency, as consistent with the experiments.
\red{We have also ignored any phase difference between the drives applied to S1 and P3, which could arise due to different capacitances on different arms of the quantum dot circuit.
This seems justified, to a first approximation, because the drive responses appear 180$^\circ$ out of phase; however, we would generally expect such a phase difference to suppress the overall magnitude of the P3 response.}

The key to deriving an effective curvature Hamiltonian is the following adiabatic approximation: when $\hbar\omega_r\ll 2t_c$, the qubit responds adiabatically to the Hamiltonian driving terms (including the cavity coupling $g_0$).
In other words, the driving terms are unable to excite the qubit.
In this adiabatic regime (which is valid for our experiments), the slowly varying terms may be absorbed into the qubit Hamiltonian, giving
\begin{equation}
H=(\tilde\varepsilon/2)\sigma_z
+t_c\sigma_x
+2\varepsilon_r\cos(\omega_rt)(a+a^\dagger)
+\hbar\omega_ra^\dagger a,
\end{equation}
where
\begin{equation}
\tilde\varepsilon=\varepsilon_0+\varepsilon_q\cos(\omega_rt)+2\hbar g_0(a+a^\dagger) .
\end{equation}
With regards to qubit dynamics, $\tilde\varepsilon$ may simply be viewed as a resonator-dependent `constant.' 
Diagonalizing within the qubit subspace then gives
\begin{equation}
H'=(E_q/2)\tilde\sigma_z
+2\varepsilon_r\cos(\omega_rt)(a+a^\dagger)
+\hbar\omega_ra^\dagger a, \label{eq:Hpcurve}
\end{equation}
where 
\begin{equation}
E_q=\sqrt{4t_c^2+\left[\varepsilon_0+\varepsilon_q\cos(\omega_rt)+2\hbar g_0(a+a^\dagger)\right]^2} , \label{eq:diagEq}
\end{equation}
and $\tilde\sigma_z$ is a Pauli matrix in the qubit frame.

We now further assume that the experiments operate in the weak-driving regime, such that $\varepsilon_q,\hbar g_0\ll t_c$.
Taylor expanding Hamiltonian (\ref{eq:Hpcurve}) up to order ${\cal O}[g_0^2,g_0\varepsilon_q]$ then gives
\begin{multline}
H'\approx (E_{q0}/2)\tilde\sigma_z
+\hbar \left[g_\|^\text{st}+g_\|^\text{dy}\cos(\omega_r t)\right]\tilde\sigma_z\left(a+a^\dagger\right) \\
+(\hbar\delta\omega/2)\,\tilde\sigma_z\left(a+a^\dagger\right)^2
+2\varepsilon_r\cos(\omega_rt)\left(a+a^\dagger\right) \\
+\hbar\omega_ra^\dagger a,
\label{eq:Hpcurve2}
\end{multline}
where
\begin{gather}
E_{q0}=\sqrt{\varepsilon_0^2+4t_c^2} 
\quad\quad \text{(static qubit energy)} ,\\
g_\|^\text{st}=\frac{\varepsilon_0 g_0}{E_{q0}} 
\quad\quad \text{(static longitudinal coupling)} ,\\
g_\|^\text{dy}=\frac{4t_c^2g_0\varepsilon_q}{E_{q0}^3}
\quad\quad \text{(dynamic longitudinal coupling)} ,\\
\delta\omega=\frac{8t_c^2\hbar g_0^2}{E_{q0}^3} 
\quad\quad \text{(dispersive coupling)} .
\end{gather}
We have dropped two inconsequential terms in Eq.~(\ref{eq:Hpcurve2}).
The first is a term $\propto\varepsilon_q\cos(\omega_rt)\tilde\sigma_z$, which adiabatically modifies the qubit energy, but averages to zero in time-averaged experiments like the ones described here.
The second term is the ${\cal O}[\varepsilon_q^2]$ term in the Taylor expansion, which causes weak oscillations $\propto\varepsilon_q^2 \cos^2(\omega_rt)\tilde\sigma_z$ in the laboratory frame, and a small, constant shift of the qubit energy splitting in the rotating frame, but does not affect the results described below.

We can simplify Eq.~(\ref{eq:Hpcurve2}) by noting that `counter-rotating' terms proportional to $a^2$or ${a^\dagger}^2$ do not conserve energy and may therefore be eliminated through a rotating wave approximate, giving
\begin{multline}
H'\approx (E_{q0}/2)\tilde\sigma_z
+\hbar \left[g_\|^\text{st}+g_\|^\text{dy}\cos(\omega_r t)\right]\tilde\sigma_z\left(a+a^\dagger\right) \\
+\hbar\delta\omega\,\tilde\sigma_z\left(a^\dagger a+1/2\right)
+2\varepsilon_r\cos(\omega_rt)\left(a+a^\dagger\right) \\
+\hbar\omega_ra^\dagger a.
\label{eq:Hpcurve3}
\end{multline}
Finally, moving to the rotating frame defined by $U=\exp(-i\omega_rt\,a^\dagger a)$ and applying a rotating wave approximation, we find that
\begin{multline}
H_\rho=U^\dagger H'U-i\hbar U^\dagger\frac{d}{dt}U
\approx \frac{E_{q0}}{2}\tilde\sigma_z
+\frac{\hbar g_\|^\text{dy}}{2}\tilde\sigma_z\left(\tilde a+\tilde a^\dagger\right) \\
+\hbar\delta\omega\,\tilde\sigma_z\left(\tilde a^\dagger \tilde a+\frac{1}{2}\right)
+\varepsilon_r\left(\tilde a+\tilde a^\dagger\right) , \label{eq:Hrho}
\end{multline}
where $\tilde a^\dagger$ and $\tilde a$ are photon creation and annihilation operators in the rotating frame.
Equation~(\ref{eq:Hrho}) contains all the curvature couplings.

\setlength{\tabcolsep}{10pt}
\renewcommand{\arraystretch}{1.2}
\begin{table*}[t]
    \caption{Best-fit parameters and uncertainties obtained using the fitting procedures described in  Appendix~\ref{fitprocedure}.}\vspace{0.1in}
    \red{
    \begin{tabular}{l c c c c c}
        \toprule
        & $(c \, \varepsilon_2) \,/\, 10^{-5}$ (V\,GHz) & $(c \, \varepsilon_3) \,/\, 10^{-5}$ (V\,GHz) & $\beta_2 \,/\, 10^{-2}$ & $\beta_3 \,/\, 10^{-3}$ & $t_c$ (GHz) \\
        \colrule
        Fig.~3(c,d) & $2.52 \pm 0.21$ & $12.93 \pm 0.60$ & $1.27 \pm 0.11$ & $2.89 \pm 0.14$ & $6.14 \pm 0.21$ \\
        Fig.~4(a)(i)    & -- & $0.92 \pm 0.02$ & -- & $3.09 \pm 0.04$ & $5.57 \pm 1.34$ \\
        Fig.~4(a)(ii)   & -- & $4.84 \pm 0.06$ & -- & $3.09 \pm 0.04$ & $5.24 \pm 0.26$ \\
        Fig.~4(a)(iii)  & -- & $12.81 \pm 0.17$ & -- & $3.09 \pm 0.04$ & $5.90 \pm 0.14$ \\
        Fig.~4(d)(iv)   & -- & $12.77 \pm 0.00$ & -- & $2.93 \pm 0.02$ & $4.65 \pm 0.08$ \\
        Fig.~4(d)(v)    & -- & $12.66 \pm 0.00$ & -- & $2.93 \pm 0.02$ & $5.51 \pm 0.09$ \\
        Fig.~4(d)(vi)   & -- & $12.57 \pm 0.00$ & -- & $2.93 \pm 0.02$ & $6.55 \pm 0.15$ \\
        \botrule
    \end{tabular}
    }
    \label{tab:raw_fit_params}
\end{table*}

As noted in the main text, the definition of the dispersive coupling $\delta\omega$ used in the preceding equations differs from the conventional definition of the dispersive coupling $\chi$ by a factor of two.
To understand this, we note that Eq.~(\ref{eq:Hrho}) can also be solved using a standard Schrieffer-Wolff perturbation method, yielding
\begin{equation}
\delta\omega \approx g_\perp^2\left(\frac{1}{\omega_q-\omega_r}+\frac{1}{\omega_q+\omega_r}\right) . 
\label{eq:Schrieffer}
\end{equation}
\red{In the conventional dispersive regime, where $|\omega_q-\omega_r|\ll \omega_r$, the second term in Eq.~(\ref{eq:Schrieffer}) is counterrotating (small), and is therefore dropped in many calculations, yielding the conventional dispersive result, $\delta\omega\approx g_\perp^2/\Delta=\chi$, where $\Delta=|\omega_q-\omega_r|$.
However in the ultra-dispersive regime, $\omega_r\ll\omega_q$, both terms have the same magnitude, and we obtain $\delta\omega\approx 2g_\perp^2/\omega_q\approx2\chi$.}

\section{IQ equations}
Equation~(\ref{eq:Hrho}) is now used to calculate the dynamics of the density matrix.
Expressing the photonic terms of the density matrix in the coherent-state basis $\{\ket{\tilde\alpha}\}$, where $a\ket{\tilde\alpha}=\tilde\alpha\ket{\tilde\alpha}$, yields the following dynamical equations~\cite{Ruskov:2019p245306}
\begin{equation}
\frac{d\tilde\alpha_\pm}{dt}=-i\left(\pm \delta\omega-i\frac{\kappa}{2}\right)\tilde\alpha_\pm-i\left(\frac{\varepsilon_r}{\hbar}\pm \frac{g_\|^\text{dy}}{2}\right) , \label{eq:alphapm}
\end{equation}
where the tilde notation ($\tilde\alpha_\pm$) indicates that we are working in the rotating frame.
Here, the $\pm$ sign corresponds to the eigenvalues of $\tilde\sigma_z$ in Eq.~(\ref{eq:Hrho}); the cavity evolution therefore depends on the state of the qubit.
Note that the first brackets in Eq.~(\ref{eq:alphapm}) correspond to diagonal terms in Eq.~(\ref{eq:Hrho}), with respect to the photon operators, while the second brackets correspond to off-diagonal terms.
Also note that we have introduced an energy-nonconserving term with strength $\kappa$ to account for photon loss from the cavity, at a rate $\kappa/2\pi$.
In the transmission experiments described in the main text, we probe the stationary states of the cavity ($\tilde\alpha_\pm^\text{st}$), which can be obtained from Eq.~(\ref{eq:alphapm}) by setting $d\tilde\alpha_\pm/dt=0$, giving
\begin{equation} \label{eq:alpha_st}
\tilde\alpha_\pm^\text{st}=-\frac{(\varepsilon_r/\hbar)\pm(g_\|^\text{dy}/2)}{\pm \delta\omega-i\kappa/2} .
\end{equation}

\setlength{\tabcolsep}{7pt}
\renewcommand{\arraystretch}{1.2}
\begin{table*}[t]
    \caption{Device parameters obtained using the methods described in Appendix~\ref{deviceparams}.}\vspace{0.1in}
    \red{
    \begin{tabular}{l c c c c c c c}
        \toprule
        & $\alpha_{\mathrm{P2}, \varepsilon}$~(eV/V) & $\alpha_{\mathrm{P3}, \varepsilon}$~(eV/V) & $\alpha_{\mathrm{S1}, \varepsilon}$~(eV/V) & $Z_{0,r}$~($\Omega$) & $\omega_r/2\pi$~(GHz) & $\kappa/2\pi$~(kHz) & $g_0/2\pi$~(MHz) \\
        \colrule
        Fig.~3(c,d) & 0.11 & 0.09 & 0.04 & 575 & 1.3038 & 124.5 & 5.5 \\
        Fig.~4(a,b) & 0.11 & 0.09 & 0.04 & 575 & 1.3038 & 124.5 & 5.5 \\
        Fig.~4(c,d) & 0.10 & 0.10 & 0.03 & 575 & 1.3038 & 124.5 & 4.1 \\
        \botrule
    \end{tabular}
    }
    \label{tab:fit_params}
\end{table*}

The coherent state amplitudes $\tilde\alpha_\pm$ are proportional to the output (transmitted) amplitude of the resonator $\tilde A$, which can be measured using a homodyne circuit.
The technique converts the rapidly oscillating microwave transmission field to a dc signal.
In brief, a copy of the ac input drive (applied to gates S1 or P3) is multiplied by the output signal from the cavity, using an IQ mixer.
A second copy of the input signal, which is phase-shifted by $\pi/2$, is also multiplied by the output signal.
Time averaging these separate products provides measurement of the $I$ and $Q$ quadratures of the cavity transmission.
The cavity input fields in this calculation are defined as
\begin{equation}
a_\text{in}^I=A\cos(\omega_r t)
\quad\quad\text{and}\quad\quad
a_\text{in}^Q=-A\sin(\omega_r t) ,
\end{equation}
and the cavity output field is defined as 
\begin{equation}
a_\text{out}=\tilde A\cos(\omega_r t+\phi_\text{out}) ,
\end{equation}
where $\phi_\text{out}$ represents the phase shift of the transmitted signal.
We note that the input field strength $A$ differs for drives applied to gates S1 vs.\ P3 \red{at a given microwave generator power}, due to their distinct circuit paths and microwave attenuations.
The product signals are given by
\begin{gather}
I=\overline{a_\text{in}^I a_\text{out}}=\frac{A\tilde A}{2}\cos(\phi_\text{out}), \\
Q=\overline{a_\text{in}^Q a_\text{out}}=\frac{A\tilde A}{2}\sin(\phi_\text{out}) ,
\end{gather}
where the bar notation denotes time averaging.
The combined IQ output is then given by
\begin{equation}
\text{IQ}_\pm\equiv\sqrt{I^2+Q^2}=c|\alpha_\pm|
=c\frac{(\varepsilon_r/\hbar)\pm(g_\|^\text{dy}/2)}{\sqrt{\delta\omega^2+\kappa^2/4}} , \label{eq:IQpm}
\end{equation}
where the prefactor $c$ accounts for the details of the measurement and room-temperature circuitry, including attenuation, signal amplification, and other effects.
This factor prevents us from obtaining an absolute estimate of the various coupling terms directly from $\text{IQ}_\pm$.
\red{For interpreting the measurements in Figs.~4(a) and 4(b) of the main text, we point out that $c \propto A \propto \varepsilon_q$, which can be seen by combining Eq.~(\ref{eq:IQpm}) with the fact that $|\alpha_\pm| \propto \tilde{A}$.}
Finally we note that the qubit remains in its ground state in all of the experiments described here, so we only need to consider the minus sign in Eq.~(\ref{eq:IQpm}).

\section{Fitting Procedure} \label{fitprocedure}

\red{In this Appendix, we explain the fitting procedure related to Eq.~(\ref{eq:IQpm}).
We note that a dc offset in the homodyne IQ output, which was not calibrated away during experiments, has been subtracted from all data presented in this work.}

\setlength{\tabcolsep}{10pt}
\renewcommand{\arraystretch}{1.2}
\begin{table*}[t]
    \caption{Upper-bound estimates for the ac detuning drive amplitude and cavity photon number computed in Appendix~\ref{amplitude_photon}.} \vspace{0.1in}
    \red{
    \begin{tabular}{l c c c c c}
        \toprule
        & Driving gate & $P_\mathrm{in}$~(pW) & $\varepsilon_q/h$~(MHz) & $\varepsilon_r/h$~(kHz) & $\langle n \rangle$ \\
        \colrule
        Fig.~3(a,c) & S1    & 20        & 27                & 340           & 30 \\
        Fig.~3(b,d) & P3    & 20        & $-137$            & 397           & 41 \\
        Fig.~4(a,b) & P3    & 2 to 20   & $-39$ to $-137$   & 119 to 424    & 4 to 46 \\
        Fig.~4(c,d) & P3    & 20        & $-152$            & 447           & 51 \\
        \botrule
    \end{tabular}
    }
    \label{tab:drivepwr_photon}
\end{table*}

\subsection{Figure 3 fits} \label{fig3fits}

The data in Fig.~3 of the main text are fit to Eq.~(\ref{eq:IQpm}).
For the case of S1 driving, we set $\varepsilon_3=0$, obtaining
\begin{equation}
\text{IQ}_-=c\,\varepsilon_2
\frac{(\beta_2/\hbar)-(2t_c^2g_0/E_{q0}^3)}{\sqrt{\delta\omega^2+\kappa^2/4}} \quad\quad\text{(S1 drive)},
\label{eq:IQS1}
\end{equation}
while for P3 driving, we set $\varepsilon_2=0$, obtaining
\begin{equation}
\text{IQ}_-=c\,\varepsilon_3
\frac{(\beta_3/\hbar)+(1-\beta_3/\beta_2)(2t_c^2g_0/E_{q0}^3)}{\sqrt{\delta\omega^2+\kappa^2/4}} \quad\quad\text{(P3 drive)}.
\label{eq:IQP3}
\end{equation}
In these equations, we note that $E_{q0}=E_{q0}(\varepsilon_0,t_c)$ and $\delta\omega=\delta\omega(\varepsilon_0,t_c,g_0)$. Since the parameters $g_0/2\pi=5.5$~MHz and $\kappa/2\pi=125$~kHz are determined independently using methods described in Appendix~\ref{deviceparams} below, and the data are obtained as a function of $\varepsilon_0$, the resulting fitting parameters are given by $(c\,\varepsilon_2)$, $(c\,\varepsilon_3)$, $\beta_2$, $\beta_3$, and $t_c$. 

The data in Figs.~3(c) and 3(d) are fit simultaneously (i.e., both datasets are fit with a single call to the fitting function) to Eqs.~(\ref{eq:IQS1}) and (\ref{eq:IQP3}), respectively. This ensures that the common fitting parameters in both equations, $\beta_2$ and $t_c$, are optimized between the two datasets. 
The best-fit parameters and uncertainties from this fitting procedure are reported in Table~\ref{tab:raw_fit_params}, and are later used in calculating the Hamiltonian terms below, in Table~\ref{tab:symbol_list}. 

After fitting, we compute and subtract the asymptotic value $\text{IQ}_-(\varepsilon_0 \rightarrow \infty) = 2 c \varepsilon_r / \hbar \kappa$ from both line cuts in Fig.~3. This value varies slightly ($\sim$10\%) between the two drive configurations, most likely due to a small difference in attenuation on the drive lines.
Since the parameter $c$ cannot be independently determined, the line cuts in Fig.~3(d) are rescaled to have the peak value 1, while the line cuts in Fig.~3(c) are scaled by the same factor.
The parameters $(c\,\varepsilon_2)$ and $(c\,\varepsilon_3)$ have  effectively arbitrary units, although we report them here in units of volts $\times$ gigahertz, corresponding to the dc output voltages from the homodyne measurements.

\subsection{Figure 4 fits} \label{fig4fits}

\red{
The IQ transmission peaks shown in Fig.~4(a) of the main text [and the larger set of peaks leading to Fig.~4(b)] are fit to Eq.~(\ref{eq:IQP3}) at each value of the ac driving amplitude $\varepsilon_q$.
Our fitting procedure is as follows.
The fitting parameter $\beta_3$ is held fixed across all peaks, since we do not expect its value to change as a function of $\varepsilon_q$.
We also set $\beta_2$ to the value of $1.27 \times 10^{-2}$ extracted from the Fig.~3 fits, because we are unable to obtain a precise estimate of this quantity without an S1-driven dataset to fit simultaneously.
The parameters $g_0/2\pi=5.5$~MHz and $\kappa/2\pi=125$~kHz are determined independently using methods described in Appendix~\ref{deviceparams}, below.
This leaves the product $c\varepsilon_3$ and $t_c$ as the remaining fitting parameters for each peak.
After fitting, we compute and subtract the asymptotic background level \hbox{$\text{IQ}_-(\varepsilon_0 \rightarrow \infty) \propto c \varepsilon_q$} from the traces in Fig.~4(a).
}

\red{
The peaks shown in Fig.~4(c) of the main text [and the larger set of peaks leading to Fig.~4(d)] are fit to Eq.~(\ref{eq:IQP3}) at each value of barrier voltage $V_\text{B3}$ using a similar procedure.
Once again, we hold $\beta_3$ fixed across all fits and set $\beta_2 = 1.27 \times 10^{-2}$; $c\varepsilon_3$ and $t_c$ are again the remaining fitting parameters for each peak.
However, in this case we have $g_0/2\pi=4.1$~MHz, due to a different device tuning for these measurements.
}
After fitting, we again determine and subtract the asymptotic background level $\text{IQ}_-(\varepsilon_0 \rightarrow \infty)$ from the traces in Fig.~\ref{fgr:tunnelcoupling}(d). 
The small ($\sim$1\%) variation observed in the background is due to time variation in the electrostatic environment and/or change in the lever arm $\alpha_\mathrm{P3,\varepsilon}$ as a function of $V_\text{B3}$.


The best-fit parameters and uncertainties for Figs.~\ref{fgr:tunnelcoupling}(a) and \ref{fgr:tunnelcoupling}(d) are reported in Table~\ref{tab:raw_fit_params}, and corresponding Hamiltonian terms are reported in Table~\ref{tab:symbol_list}.

Peak amplitudes plotted in Fig.~4(c) are extracted by applying Lorentzian fits to the data; these amplitudes are also shown as dark teal circles in Fig.~\ref{fig:tc_vb3}(a) below. 
The arbitrary units in Figs.~4(c) and 4(d) are scaled so that the tallest peak (labeled iv) has amplitude 1.

\section{Device parameter measurements} \label{deviceparams}

\subsection{Gate lever arms} \label{leverarms}

We extract gate-to-dot lever arms by measuring thermal broadening of electron transitions~\cite{Gabelli:2006p499}. At finite temperature, thermal fluctuations of the chemical potential lead to charge transitions of finite energy width, which we measure by sweeping the dc plunger voltage $V_\mathrm{P3}$ through a dot 3 charging transition, which occurs at the voltage $V_0$.
In the limit $k_B T_e \gg \hbar \omega_r$, the transition linewidth is thermally limited, and the resonator's response to electron tunneling should follow the derivative of the Fermi-Dirac distribution,
\begin{equation}
    |\text{IQ}| = -|A| \sech^2{\bigg[\frac{\alpha_\mathrm{P3,3}(V_\mathrm{P3} - V_0)}{2 k_B T_e}\bigg]} + B,
    \label{eq:IQalpha}
\end{equation}
where $k_B$ is the Boltzmann constant, $T_e$ is the electron temperature, $A$ is the resonator suppression amplitude, $B$ is the blockade transmission signal, and $\alpha_\mathrm{P3,3}$ is the P3-to-dot 3 lever arm \cite{Kouwenhoven:1997p1384}. As the cryostat temperature is varied, the electron temperature has been observed to follow the trend
\begin{equation}
    T_e = \sqrt{T_{\mathrm{mc}}^2 + T_{e0}^2},
    \label{eq:Tealpha}
\end{equation}
where $T_{\mathrm{mc}}$ is the cryostat mixing chamber temperature and $T_{e0}$ is the base electron temperature when $T_{\mathrm{mc}}$ is extrapolated to zero~\cite{Simmons:2011p156804}.

\begin{figure}[b]
    \centering
    \includegraphics[width=6cm]{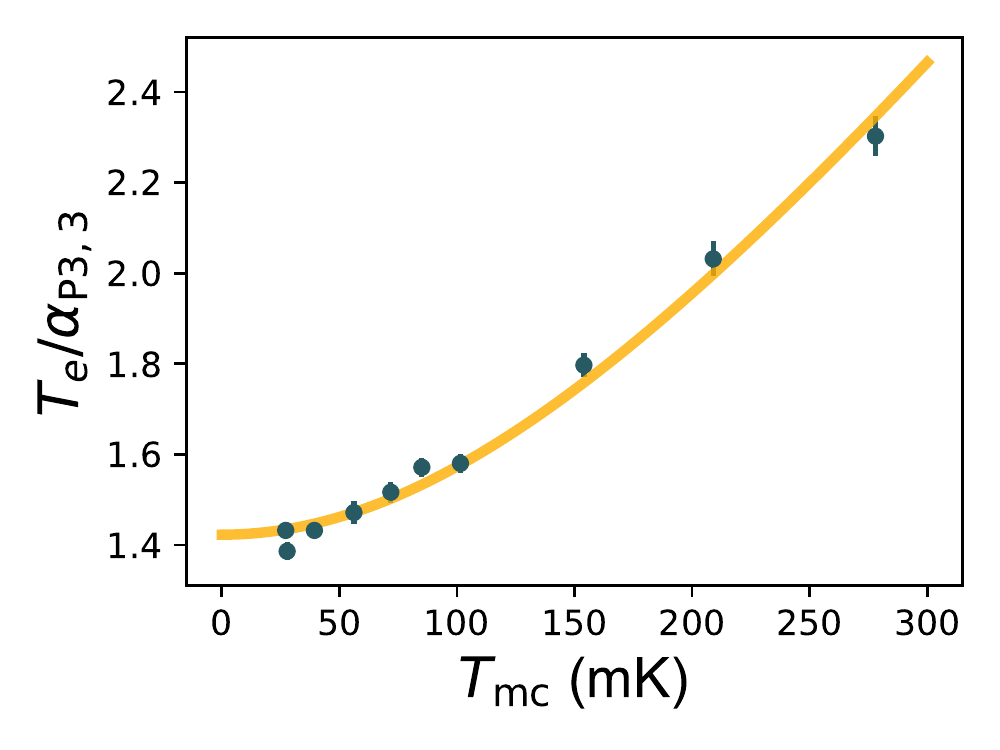}
    \caption{Data obtained from thermal broadening measurements for $\alpha_{\mathrm{P3},3}$. Data points extracted from charge-transition line fits are plotted against the mixing chamber temperature. Fitting these data to Eq.~(\ref{eq:Tealpha}) yields $\alpha_{\mathrm{P3},3} = 0.149$~eV/V and the base electron temperature $T_{e0} = 212$~mK.}
    \label{fig:alphaplot}
\end{figure}

Working at a device tuning comparable to the one used in measurements for Figs.~3 and 4(d)(vi), we measure reservoir transition widths at several cryostat temperatures. We then extract a base electron temperature $T_{e0}~=~212$~mK and a P3-to-dot 3 lever arm $\alpha_{\mathrm{P3},3} = 0.149$~eV/V by fitting to Eqs.~(\ref{eq:IQalpha}) and (\ref{eq:Tealpha}). The electron temperature extracted from the fitting is well described by Eq.~(\ref{eq:Tealpha}), as demonstrated in Fig.~\ref{fig:alphaplot}.

The other plunger lever arms are calculated from the slopes of transition lines in stability diagrams where both plungers are swept:
\begin{equation} \label{eq:alphaP3ep}
    \alpha_{\mathrm{P3}, \varepsilon} = \alpha_{\mathrm{P3}, 3} - \alpha_{\mathrm{P3}, 2} = \alpha_{\mathrm{P3}, 3} \frac{m_\mathrm{pol}}{m_3} \bigg( \frac{m_3 - m_2}{m_2 + m_\mathrm{pol}} \bigg),
\end{equation}
\begin{equation} \label{eq:alphaP2ep}
    \alpha_{\mathrm{P2}, \varepsilon} = \alpha_{\mathrm{P2}, 2} - \alpha_{\mathrm{P2}, 3} = \frac{\alpha_{\mathrm{P3},\varepsilon}}{m_\mathrm{pol}},
\end{equation}
\begin{equation} \label{eq:alphaP22}
    \alpha_{\mathrm{P2},2} = \frac{m_3 \alpha_{\mathrm{P2},\varepsilon} + \alpha_{\mathrm{P3},\varepsilon}}{m_3 - m_2},
\end{equation}
where \hbox{$m_i = (\Delta V_\mathrm{P2} / \Delta V_\mathrm{P3})_i$} are the slopes of dot 2 transition lines, dot 3 transition lines, or polarization lines in $V_\mathrm{P2}$-$V_\mathrm{P3}$ space. Similarly, the detuning lever arm for gate S1 is estimated using
\begin{equation}
    \alpha_{\mathrm{S1}, \varepsilon} = \alpha_{\mathrm{S1}, 2} - \alpha_{\mathrm{S1}, 3} = \alpha_{\mathrm{P2}, 2}\bigg(\frac{\Delta V_\mathrm{P2}}{\Delta V_\mathrm{S1}} \bigg)_2 - \alpha_{\mathrm{P3}, 3}\bigg(\frac{\Delta V_\mathrm{P3}}{\Delta V_\mathrm{S1}} \bigg)_3,
    \label{eq:alphaS1ep}
\end{equation}
where S1 lever arms are calculated from the slopes of transition lines in two-dimensional scans in which both gates are swept.

Lever arms used for each fit are reported in Table~\ref{tab:fit_params}. We have measured $\alpha_{\mathrm{P2},2}$ and $\alpha_{\mathrm{P3},3}$ at several additional device tunings using both thermal broadening and electron transport bias spectroscopy techniques in order to further verify that the above method provides reasonable estimates.

\begin{figure*}[t]
    \centering
    \includegraphics[width=1.0\textwidth]{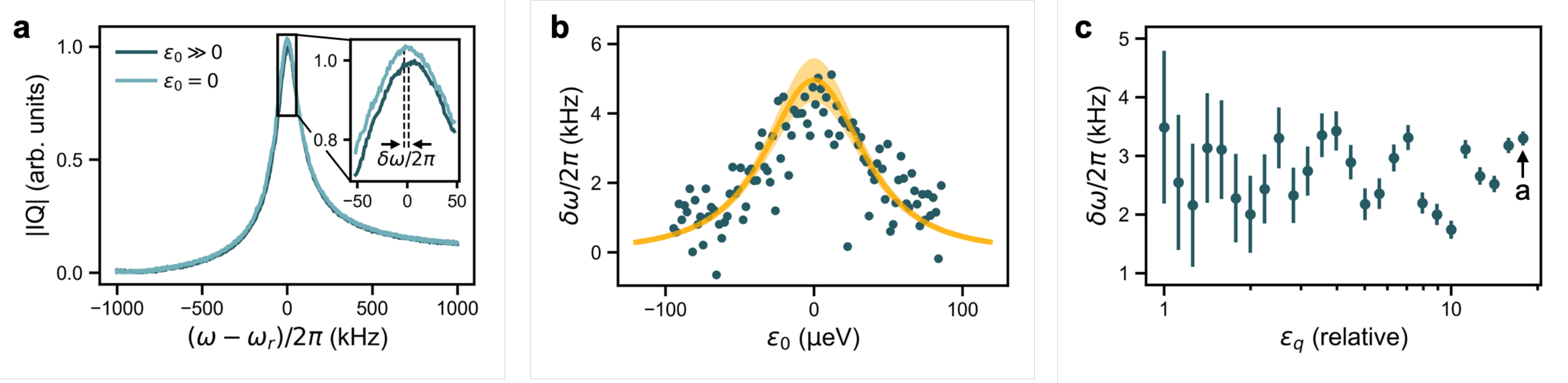}
    \caption{
        \red{
        Direct measurements of the dispersive-like coupling from cavity resonance shifts, for an ac drive applied to gate P3.
        (a) Cavity transmission as a function of driving frequency $\omega/2\pi$, measured at the interdot charging transition ($\varepsilon_0 = 0$, lighter blue), and deep within the Coulomb blockade regime ($\varepsilon_0 \gg 0$, dark blue). Inset: zoomed-in view showing the small shift of the resonant peak due to the dispersive-like coupling $\delta\omega$.
	(b) Resonant peak shifts are obtained as a function of $\varepsilon_0$. The dark-yellow curve and light-yellow $1\sigma$ confidence interval show the prediction for $\delta\omega$, obtained from Eq.~(A12), using fitting results for $t_c$ obtained in Fig.~3(d) of the main text.
        (c) Directly measured resonant frequency shifts plotted as a function of ac driving amplitude $\varepsilon_q$, in relative units. The data point corresponding to (a) is indicated with an arrow.        }    }
    \label{fig:deltaomega}
\end{figure*}

\subsection{Qubit-cavity bare coupling} \label{barecoupling}

The bare qubit-cavity coupling $g_0$ is calculated from device parameters according to the equation \cite{Childress:2004p042302}
\begin{equation}
    g_0 = \frac{\alpha_{\mathrm{S1}, \varepsilon} \, \omega_r}{2 e} \sqrt{\frac{2Z_{0,r}}{h/e^2}}.
    \label{eq:g0}
\end{equation}
Here, the characteristic impedance of the resonator, $Z_{0,r} = 575~\Omega$, is estimated from a quasi-static impedance analysis performed using COMSOL Multiphysics \cite{Holman:2021p137}.
The values of $Z_{0,r}$ and $g_0$ used in each fit are reported in Table~\ref{tab:fit_params}.

\subsection{Microwave resonator parameters} \label{cavitydecay}

The cavity resonant frequency $\omega_r$ and loaded quality factor $Q_L$ are extracted from resonance transmission curves measured with a vector network analyzer. Complex-plane fits to the curves yield $Q_L = 10,470 \pm 32$ for S1-drive and $Q_L = 10,476 \pm 46$ for P3-drive; the two values are very similar and we use their average in our analysis. The cavity decay rate $\kappa = \omega_r / Q_L$ can then be calculated. Both $\omega_r$ and $\kappa$ are reported in Table~\ref{tab:fit_params}.

\section{Microwave drive energy and cavity photon number} \label{amplitude_photon}

\subsection{Drive energy upper bound}

While the analysis method used in the main text does not provide a precise measure of the qubit detuning drive amplitude $\varepsilon_q$, we can obtain a rough estimate by considering the microwave power applied in experiments.
\red{
For the measurements reported in Fig.~3 of the main text, a microwave generator is used to apply a driving power of 6~dBm to the measurement circuit, which is configured with approximately 33~dB of attenuation at room temperature.
The measurements in Figs.~4(a) and 4(b), use $-5$ to 6~dBm generator power and 33~dB room-temperature attenuation; those in Figs.~4(c) and 4(d) use 15~dBm of generated power and have 42~dB of room-temperature attenuation.}
For all measurements, an additional 40~dB of attenuation is incurred inside the cryostat.
Furthermore, the quantum dot device is engineered with buried coplanar waveguide bias leads whose characteristic impedance is $Z_\mathrm{0,g} \approx 1~\Omega$ \cite{Holman:2021p137}.
Microwave attenuation at the bias leads has not been fully characterized; however, based on SPICE simulations of our device, we expect a minimum attenuation of 10~dB due to impedance mismatch between the leads and the $50~\Omega$ coaxial cabling inside the cryostat.
Summing the above contributions gives an estimated input power of $P_\mathrm{in} \lesssim -77$~dBm~=~20~pW for all measurements.

When ac modulation is applied to plunger gate P3, the detuning drive amplitude is related to the input power by
\begin{equation} \label{eq:epsilon_q}
    \varepsilon_q \approx \varepsilon_3 = -\alpha_\mathrm{P3,\varepsilon} \sqrt{2 Z_\mathrm{0,g} P_\mathrm{in}},
\end{equation}
where $\alpha_\mathrm{P3,\varepsilon}$ is the detuning lever arm, and the factor of $\sqrt{2}$ converts between rms and peak voltage ampitudes.
Note that for this estimate, we ignore the small crosstalk correction proportional to $(\beta_3 / \beta_2 )$ seen in Eq.~(A1).
Equation~(\ref{eq:epsilon_q}) is not applicable for the measurements reported in Figs.~3(a) and 3(c).
In those experiments, the gate S1 is driven through the $\lambda/4$ resonator segment, resulting in additional voltage attenuation.
We estimate the resulting $\varepsilon_q$ values by multiplying the output of Eq.~(\ref{eq:epsilon_q}) by the ratio of best-fit parameters $(c\,\varepsilon_2)/(c\,\varepsilon_3)$ reported in Table~\ref{tab:raw_fit_params}.
Thus for all experiments performed here, we estimate drive amplitudes of $|\varepsilon_q/h| \lesssim 152$~MHz, obtaining the full set of results listed in Table~\ref{tab:drivepwr_photon}. 
These drive strengths are comparable to those used in similar qubit experiments \cite{Corrigan:2021p127701}.

We emphasize again that these values should be treated as approximate upper bounds due to the uncertainty in attenuation at the quantum dot bias leads.

\setlength{\tabcolsep}{10pt}
\renewcommand{\arraystretch}{1.2}
\begin{table*}[t]
\caption{Estimated energy coupling parameters for experiments reported in the main text. 
The tunnel coupling $t_c$ is obtained from fitting; dot-cavity couplings $\delta\omega$ and $g_\|^\text{dy}$ are computed according to Eqs.~(A11) and (A12) using $t_c$ and parameters from Tables~\ref{tab:fit_params} and \ref{tab:drivepwr_photon}.
Note that the values for $g_\|^\text{dy}$ are upper-bound approximations, based on the $\varepsilon_q$ values reported in Table~\ref{tab:drivepwr_photon}.
The reported values all correspond to the peaks, centered at $\varepsilon_0=0$.}\vspace{0.1in}
\red{
\begin{tabular}{l c c c c}
 \toprule
 & $t_c/h$~(GHz) & $\delta\omega/2\pi$~(kHz) & $g_\|^\text{dy}/2\pi$~(kHz, approx.) & $|g_\|^\text{dy}/\delta\omega|$\\
 \colrule
 Fig.~3(c)      & 6.1   & 4.9   & 12.1      & 2.5 \\
 Fig.~3(d)      & 6.1   & 4.9   & $-61.4$   & 12.4 \\
 Fig.~4(a)(i)   & 5.6   & 5.4   & $-19.3$   & 3.5 \\
 Fig.~4(a)(ii)  & 5.2   & 5.8   & $-45.7$   & 7.9 \\
 Fig.~4(a)(iii) & 5.9   & 5.1   & $-63.9$   & 12.4 \\
 Fig.~4(d)(iv)  & 4.6   & 3.7   & $-67.6$   & 18.4 \\
 Fig.~4(d)(v)   & 5.5   & 3.1   & $-57.0$   & 18.4 \\
 Fig.~4(d)(vi)  & 6.5   & 2.6   & $-47.9$   & 18.4 \\
 \botrule
\end{tabular}
}
\label{tab:symbol_list}
\end{table*}

\subsection{Photon number}

\begin{figure*}[t]
    \centering
    \includegraphics[width=0.6\textwidth]{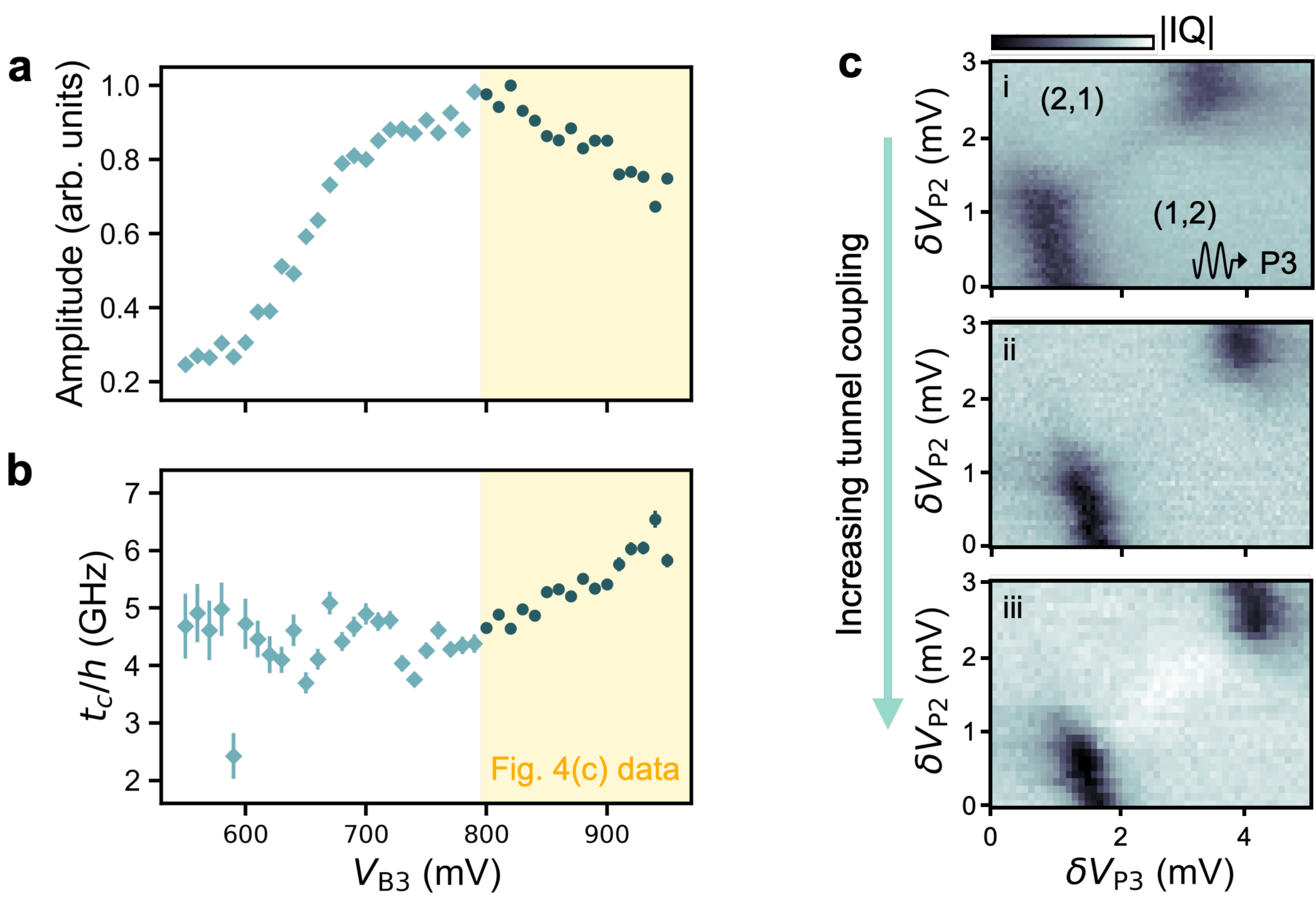}
    \caption{
    Cavity transmission responses measured for P3 driving near the interdot charging transition, extending the data set in Fig.~\ref{fgr:tunnelcoupling}(c) of the main text to lower $V_\text{B3}$ 
    values.
    (a) Cavity transmission peak amplitudes, extracted by fitting transmission peaks to Lorentzian forms as a function of dc barrier voltage $V_\text{B3}$.
    The data have been normalized so that the largest measured amplitude corresponds to 1 arbitrary unit.
    Error bars are smaller than the markers.
    (b) Tunnel coupling $t_c$ as a function of $V_\text{B3}$ estimated from transmission peak fits to Eq.~(C2).
    The low-$t_c$ outlier at $V_\text{B3} = 590$~mV is believed to be the result of electrostatic instability during the measurement of that trace.
    In both (a) and (b), the dark teal circle markers at $V_\text{B3} \geq 800$~mV correspond to data shown in Fig.~\ref{fgr:tunnelcoupling}(c) of the main text; light teal diamond markers at $V_\text{B3} < 800$~mV show an extended data set discussed in Appendix~\ref{lowVB3}.
    \red{
    (c) Three other device tunings, measured in sequence, and measured separately from the data in (a) and (b). Moving from (i) to (iii), the response changes from a dip in (i) to a peak in (iii), passing through a region of no response, (ii).
    }
    }
    \label{fig:tc_vb3}
\end{figure*}

The average number of photons populating the resonator $\langle n \rangle$ is given by the stationary value of the cavity coherent state $\tilde{\alpha}_\pm^\text{st}$. 
Squaring Eq.~(\ref{eq:alpha_st}) gives
\begin{equation} \label{eq:photon_number}
    \langle n \rangle = | \tilde{\alpha}_\pm^\text{st} |^2 \approx \frac{4 \varepsilon_r^2}{\hbar^2 \kappa^2}.
\end{equation}
Note that this quantity is dominated by the direct resonator drive, and we neglect small contributions from the couplings $g_\parallel^\text{dy}$ and $\delta\omega$. 
Using Eq.~(A2) and values from Tables~\ref{tab:raw_fit_params} and \ref{tab:drivepwr_photon} to compute $\varepsilon_r$, we estimate photon populations of \red{$\langle n \rangle \lesssim 50$} for all measurements reported here.
The full set of results are listed in Table~\ref{tab:drivepwr_photon}.

The number of thermal photons in the resonator due to finite system temperature and blackbody radiation from resistive loads is of order $\langle n \rangle \sim 1$ \cite{Carter:2019p092602}. 
This residual population \red{should be negligible for most measurements,} compared to photons produced by direct resonator drive.

We have considered the impact of higher-order corrections to the curvature couplings that may become significant at large photon number \cite{Ruskov:2019p245306}. 
For our system, we expect such corrections to be unimportant when $\langle n \rangle \lesssim 10^3$-$10^4$, due to the quantum-nondemolition nature of the curvature interactions and the fact that we work in the adiabatic coupling regime.

\section{Calculated energy couplings}

Using methods described here and in the main text, we obtain estimates for $t_c$, $g_0$, etc.
In Table~\ref{tab:symbol_list}, we report these $t_c$ values and the corresponding energy couplings, computed using Eqs.~(A11) and (A12).

\begin{figure*}[t]
    \centering
    \includegraphics[width=0.9\textwidth]{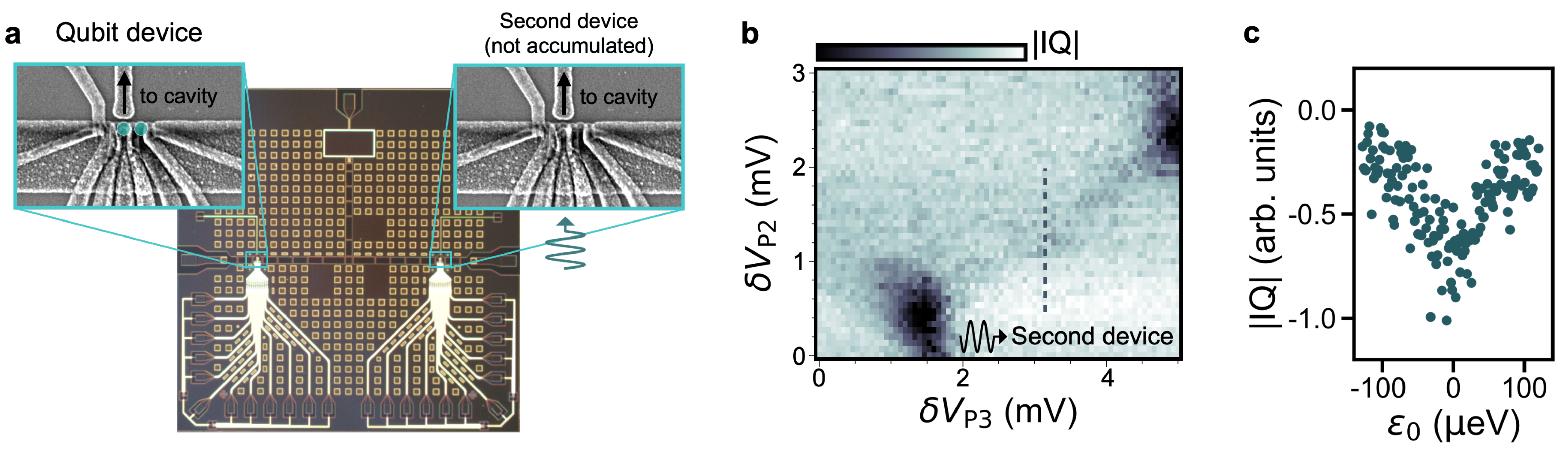}
    \caption{Cavity drive via a second on-chip device. (a) Dark-field optical micrograph of the qubit die with two quantum dot devices located in the blue boxed regions. SEM micrographs of the nominally identical devices are also shown. A 1.3~GHz microwave drive applied to the center plunger gate of the right-side device excites the resonator and can be used for charge sensing. (b) IQ transmission measurement of the double dot tuned to its (2,1)-(1,2) charge occupation transition while driving the cavity via the second device plunger. (c) Line cut across the interdot charge transition indicated in (b). Resonator transmission is suppressed at the charge degeneracy point. The scaling used for the arbitrary units here is the same used in Fig.~3.}
    \label{fig:second_device}
\end{figure*}

\red{\section{Additional data: direct measurement of dispersive-like coupling}}

\red{
For all the figures shown in the main text, the dispersive coupling $\delta\omega$ was estimated from Eq.~(A12), where $t_c$ values were obtained from the fitting analysis described in the main text, and $g_0$ was estimated using the geometrical parameters of the device.
However, $\delta\omega$ can also be measured directly from the shift of the cavity resonance frequency, when the cavity is coupled to the double dot.
Figure~7(a) shows a typical analysis, where the ac drive is applied to gate P3.
Here, the dark-blue cavity resonance peaks are obtained deep in the Coulomb blockade regime ($\varepsilon_0 \gg 0$), where the qubit-cavity coupling is tiny.
The lighter-blue curve is obtained at the interdot charging transition ($\varepsilon_0 = 0$), where the coupling is strongest.
Equation~(A12) gives a prediction for the state-dependent frequency shift, which is analogous to the effect of $\chi$ in the conventional dispersive regime.
The blown-up peaks (inset) show the frequency shift caused by the dispersive-like coupling, with $\delta\omega/2\pi \approx 3.3$~kHz.
In the ultra-dispersive limit, such shifts are found to be much smaller than the resonant linewidth.
We note that the double dot remains in its ground state for all experiments in this work, so the direction of the resonance shift caused by the coupling is always towards lower frequencies.}

\red{
This peak-shift analysis is repeated for different detuning values about the interdot charging transition, as shown in Fig.~7(b).
Here, the experimental tuning is the same tuning used for Figs.~3(b) and 3(d) of the main text.
(Note that, although the measured frequency shifts are actually negative, we plot their absolute value here, since $\delta\omega$ is defined as a positive quantity in Appendix~A.)
We note that the yellow curve shown in the figure is not a fit; rather, it shows the \emph{predicted} frequency shift obtained from Eq.~(A12), using the $t_c$ values obtained from the fit shown in Fig.~3(d) of the main text.
Here, the lighter yellow shading corresponds to the $1\sigma$ confidence interval, estimated from parameter uncertainties of the fitting algorithm.
The good agreement between the direct measurement and the theoretical prediction supports our theoretical interpretation of the results and validates our fitting methods.}

\red{Finally in Fig.~7(c), we show the experimentally measured frequency shifts as a function of the ac driving strength $\varepsilon_q$.
Here, the specific data point corresponding to Fig.~7(a) is indicated by the arrow.
While some spread is observed in the data, we observe no systematic variation of $\delta\omega$, which agrees with the conclusions of Fig.~4(b) of the main text.
We note however that the two data sets were acquired at different device tunings, and therefore are not directly related.}

\section{Additional data: low-barrier voltage behavior} \label{lowVB3}

The fitting results reported in Table~IV confirm that smaller $t_c$ values yield larger absolute values of $g_\|^\text{dy}$, as desired for strong qubit-resonator couplings. 
In this section, we explore the possibility of extending the data in Fig.~\ref{fgr:tunnelcoupling}(c) of the main text to lower $t_c$ values.

Figure~8(a) shows results of such an analysis, which includes all the data shown in Fig.~4(c), but also includes data obtained at lower values of $V_\text{B3}$, where we expect to observe lower $t_c$ values. 
Using the same IQ peak-fitting procedure as in Fig.~4(d), we also determine the corresponding $t_c$ values for this data, as reported in Fig.~8(b). 
The data reveal surprising behavior that is inconsistent with a continuous downward trend of $t_c$, in the low-$V_\text{B3}$ regime. 
Indeed, for $V_\text{B3}$ values below those reported in Fig.~4(c), the $t_c$ results remain roughly constant, but exhibit distinct nonmonotonic features. 

\begin{figure*}[t]
    \centering
    \includegraphics[width=0.5\textwidth]{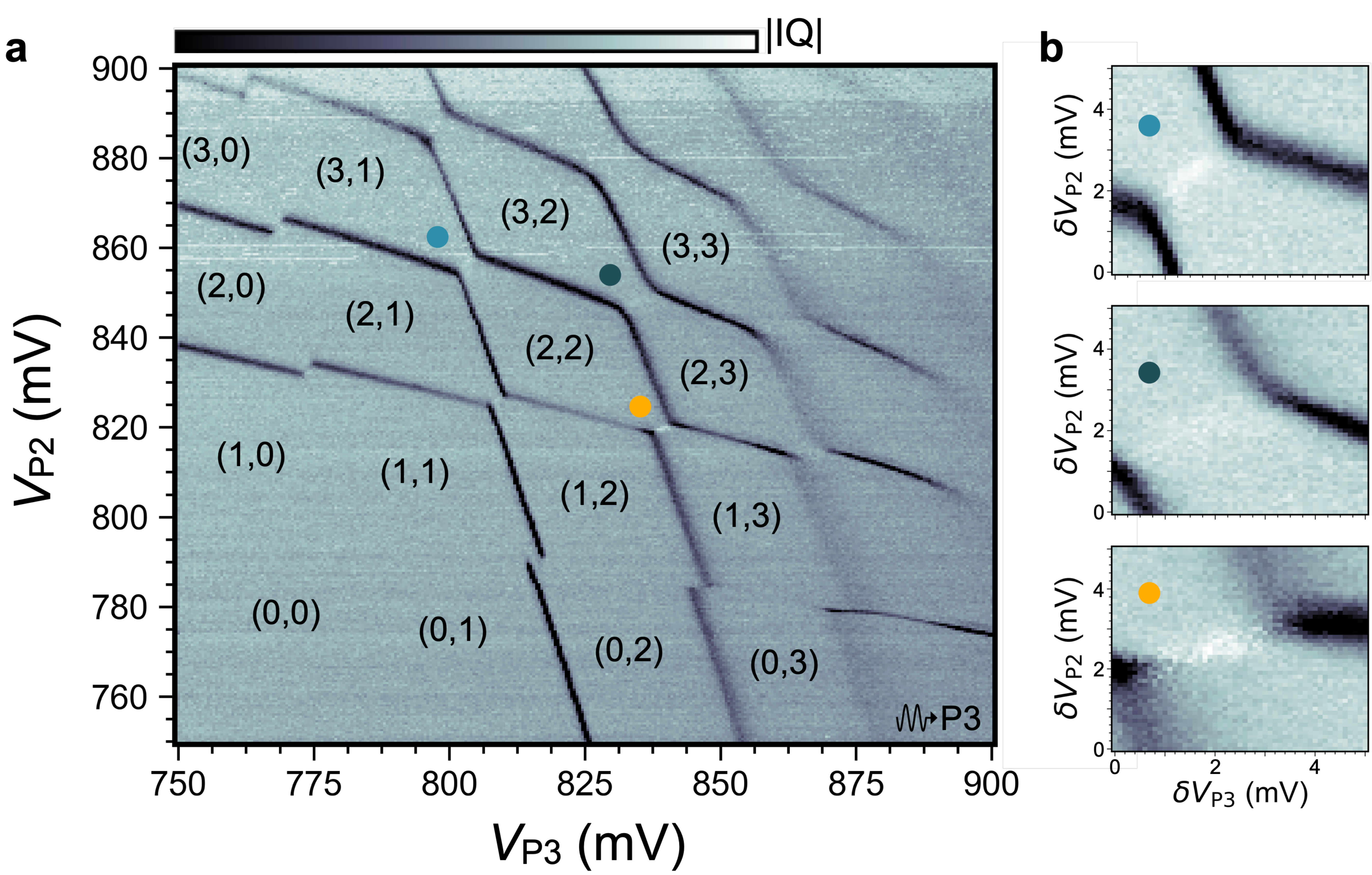}
    \caption{Enhanced transmission at several charge occupations. (a) Large-scale charge stability diagram for the double quantum dot. Experiments in the main text are performed at the (2,1)-(1,2) charge occupation. Enhanced resonator transmission is also visible at several polarization lines for the same tuning. (b) Zoomed-in stability diagrams showing enhanced resonator transmission at the polarization lines. Colored circles correspond to matching circles in (a), which indicate the voltage-space location for each scan. All plots are acquired at the same barrier and screening-gate voltage tunings. Color scale is the same as in (a).}
    \label{fig:other_occupations}
\end{figure*}

To explain this behavior, we note that lower $V_\text{B3}$ values should yield larger tunnel barriers between dots 2 and 3. Such behavior is not observed in Fig.~8(b).
Instead, the data are consistent with raising the tunnel barrier so far that the dots are squeezed out the top side of the gated region, in the device shown in Fig.~2(d) of the main text. 
This can occur here because no screening gate is present in the top portion of the device, and the dots are therefore more weakly confined in this region. 
The distinct features observed in Fig.~\ref{fig:tc_vb3}(b) are likely caused by first one, then the other of the dots being squeezed out of its intended location. 
The sudden drop of amplitude in Fig.~\ref{fig:tc_vb3}(a), at $V_\text{B3}\approx 700$~mV, correlates with the onset of the rightmost bump feature in Fig.~\ref{fig:tc_vb3}(b). 
We speculate that this occurs when dot 3 leaves its intended confinement location, resulting in a sudden reduction of the lever arm between dot 3 and gate P3; in turn, this causes a sudden drop in the driving amplitude, $\varepsilon_3$.

\red{
Figure~\ref{fig:tc_vb3}(c) shows a sequence of transmission responses measured at three additional device tunings, also in the low-$V_\text{B3}$ regime.
These scans are obtained by simultaneously varying $V_\text{B3}$ and the right-side reservoir-accumulation gate voltage [see the right-hand side of Fig.~\ref{fgr:SEM}(d)], causing $t_c$ to increase upon moving from panels (i) to (iii).
For these measurements, we also observe a sign change of the  IQ peak height.
However, in contrast with Fig.~\ref{fgr:stability} of the main text, we do not change driving ports in this case.
Interestingly, we are again able to fully extinguish the transmission response, as shown in panel~(ii), for a particular tuning.
The progression illustrates an interplay between the longitudinal ($g_\|^\text{dy}$) and dispersive ($\delta\omega$) couplings, showing that they can be independently tuned.
We speculate that the effect can be explained by lateral dot motion, which modifies both the gate lever arms and the bare coupling $g_0$.
The possible dynamics are likely more complicated than those of Figs.~\ref{fig:tc_vb3}(a) and \ref{fig:tc_vb3}(b), however, because in Fig.~\ref{fig:tc_vb3}(c), the changes in $V_\text{B3}$ were at least partially compensated by changes in the reservoir accumulation gate.
}

To summarize, reducing $V_\text{B3}$ below 800~mV in this device does not appear to cause the intended effect of lowering $t_c$. 
The IQ peak-fitting results in this regime are therefore meaningless, because the fitting equations do not include the relevant physics. 
We emphasize however, that such behavior occurs here, only because of the device geometry, and would not be expected to occur in a device with an additional, top screening gate. 
It is therefore important to explore the possibility of lowering $t_c$ values in future experiments.
\red{We also note that intentionally shifting the lateral positions of the quantum dots may provide additional means for tuning $g_\|^\text{dy}$ and $\delta\omega$ independently.}

\section{Additional data: cavity drive via a second on-chip device} \label{seconddevicedrive}

Charge sensing in the main text is performed by applying an ac drive to device gates S1 or P3. We are also able to charge sense by driving the plunger gate of a second, unaccumulated quantum dot device on the qubit die. The second device is nominally identical to the qubit device used in main-text experiments with an upper screening gate connected to the same microwave resonator, as shown in Fig.~\ref{fig:second_device}(a). The ac drive excites the cavity and modulates the S1 gate voltage. Charge stability diagrams acquired with this technique, therefore, look very similar to those acquired when driving S1, with a dip in the IQ signal at the polarization line. Figure~\ref{fig:second_device}(b) shows a stability diagram acquired at the same tuning as the Fig.~\ref{fgr:stability} measurements. The line cut in Fig.~\ref{fig:second_device}(c) shows the cavity response near the charge degeneracy point.

\section{Additional data: transmission enhancement at other charge occupations} \label{otheroccupations}

Main-text experiments are performed at the (2,1)-(1,2) charge configuration. However, we have observed resonator transmission enhancement at several additional interdot transitions when longitudinal coupling is activated via the P3 drive. The large charge stability diagram in Fig.~\ref{fig:other_occupations}(a) shows enhanced transmission for at least four transitions: (2,1)-(1,2); (3,1)-(2,2); (2,2)-(1,3); and (3,2)-(2,3). Higher-resolution scans at the latter three transitions are shown in Fig.~\ref{fig:other_occupations}(b). We note that some transitions in Fig.~\ref{fig:other_occupations}(a) appear to show suppressed transmission [e.g., (4,0)-(3,1)], while others have no visible polarization line [e.g., (1,1)-(0,2)]. We believe this variation in resonator response is strongly dependent on tunnel coupling for each configuration, as discussed in the main text. It may further depend on multi-electron states for some charge occupations; this topic lies beyond the scope of our analysis.

\section{Device fabrication and experimental setup}

Detailed information regarding fabrication of the flip-chip device and the measurement circuit used for experiments performed in this work can be found in Ref.~\cite{Holman:2021p137}.

\end{document}